\documentclass[10pt,conference,onecolumn]{IEEEtran}
\IEEEoverridecommandlockouts
\usepackage{cite}
\usepackage{amsmath,amssymb,amsfonts}
\usepackage{algorithmic}
\usepackage{graphicx}
\usepackage{textcomp}
\usepackage{xcolor}
\usepackage{subcaption}
\usepackage{stackengine} 
\usepackage{booktabs}
\usepackage[hyphens]{url}

\def\BibTeX{{\rm B\kern-.05em{\sc i\kern-.025em b}\kern-.08em
    T\kern-.1667em\lower.7ex\hbox{E}\kern-.125emX}}

\usepackage{fancyhdr}

\fancypagestyle{firstpage}{
  \fancyhf{} 

  \fancyfoot[L]{\footnotesize
    © 2025 IEEE. Personal use of this material is permitted. Permission from IEEE must be obtained for all other uses, in any current or future media, including reprinting/republishing this material for advertising or promotional purposes, creating new collective works, for resale or redistribution to servers or lists, or reuse of any copyrighted component of this work in other works.}
}

\AtBeginDocument{\thispagestyle{firstpage}}

\begin{document}

\title{
Qubit Health Analytics and Clustering for HPC-Integrated Quantum Processors
}

\author{\IEEEauthorblockN{1\textsuperscript{st} Xiaolong Deng}
\IEEEauthorblockA{
\textit{Leibniz Supercomputing Centre}\\
Garching, Germany \\
}
\and
\IEEEauthorblockN{2\textsuperscript{nd} Laura Schulz}
\IEEEauthorblockA{\textit{Argonne National Laboratory} \\
Chicago, United States \\
}
\and
\IEEEauthorblockN{3\textsuperscript{rd} Martin Schulz}
\IEEEauthorblockA{\textit{Technical University of Munich} \\
Garching, Germany \\
}

}

\maketitle
\thispagestyle{firstpage}

\begin{abstract}
Quantum computing in supercomputing centers requires robust tools to analyze calibration datasets, predict hardware performance, and optimize operational workflows. This paper presents a data-driven framework for processing calibration metrics. Our model is based on a real calibration quality metrics dataset from our in-house 20-qubit NISQ device and for more than 250 days. We apply detailed data analysis to uncover temporal patterns and cross-metric correlations. Using unsupervised clustering, we identify stable and noisy qubits. We also validate our model using GHZ state experiments. Our study provides health indicators as well as hardware-driven maintenance and recalibration recommendations, thus motivating the integration of relevant schedulers with HPCQC workflows.

\end{abstract}


\section{Introduction}\label{sec:intro}
Quantum computing promises to solve complex problems in optimization, materials simulation and cryptography that are intractable for classical systems\cite{nielsen00,RevModPhys.94.015004,Dalzell_McArdle_Berta}. However, today's noisy intermediate-scale quantum (NISQ) devices\cite{preskill2018quantum} exhibit significant variability due to environmental noise, hardware imperfections and microscopic defects such as two-level systems (TLS). To maintain stable and high-fidelity operation, superconducting quantum processors require frequent calibration, adjusting parameters such as qubit and resonator frequencies, pulse amplitudes and shapes, and tunable couplings, to compensate for drift and instability\cite{10.1063/1.5089550, arute2019quantum}.

Calibration quality metrics, including coherence time $T_1$ and $T_2$, readout fidelity and gate fidelity serve as key indicators of device performance. Fluctuations or drops in these metrics signal the need for recalibration or may point to underlying hardware issues. While a full calibration tunes all parameters from scratch, in practice, it is neither efficient nor practical. Typically daily operation relies on incremental or adaptive updates based on the parameters of previous cycles, such that only a small part or limited range of parameters are fine-tuned and optimized. This greatly reduces overhead and maximizes system availability. And drastic fluctuations might also be avoided.

Traditional calibration management often relies on manual inspection and simple threshold-based alerts. As quantum systems scale toward hundreds or thousands of qubits, this becomes infeasible. The volume and complexity of calibration data grows rapidly, and subtle correlations or emerging failure modes are difficult to be detected. In supercomputing centers, where hybrid classical-quantum workflows are increasingly integrated into high-performance computing (HPC) environments\cite{britt2017high, humble2021quantum, 10007772, elsharkawy2023}, intelligent calibration management becomes critical for optimizing circuit mapping, execution success rates, and maintenance planning\cite{10.1145/3297858.3304075, 8806892, tannu2019not, 9349092, 9951287, 9773202, 10313739, bhoumik2024resourceawareschedulingmultiplequantum, jeng2025modularcompilationquantumchiplet, mahesh2025conqurecoexecutionenvironmentquantum}. Without timely and intelligent recalibration\cite{10128756}, circuit execution can fail more often, resource utilization may decrease, and application performance suffers.

Recent research has focused on the variability and drift of superconducting qubits, automated calibration and quantum device health monitoring. Prior work has addressed crosstalk detection\cite{Sarovar2020detectingcrosstalk, PhysRevLett.131.210802}, graph-based models for calibration dependencies\cite{kelly2018physicalqubitcalibrationdirected}, and open source calibration frameworks\cite{pasquale2024qibocalopensourceframeworkcalibration, qua-platform_qualibrate_2025}. Machine learning is increasingly used for quantum circuit compilation\cite{9384317}, quantum error correction\cite{alphaqubit}, quantum device characterization and optimization\cite{alexeev2024artificialintelligencequantumcomputing,acampora2025quantumcomputingartificialintelligence}, from anomaly detection\cite{PhysRevResearch.6.033329} to predictive modeling\cite{Mavadia2017Prediction, PhysRevApplied.9.064042}, with growing interest in unsupervised clustering and graph analytics for circuit mapping and scheduling\cite{10.1145/3297858.3304075, tannu2019not}.

\begin{figure}[ht]
    \centering
    \hspace{1.5cm}
    \includegraphics[width=0.6\linewidth]{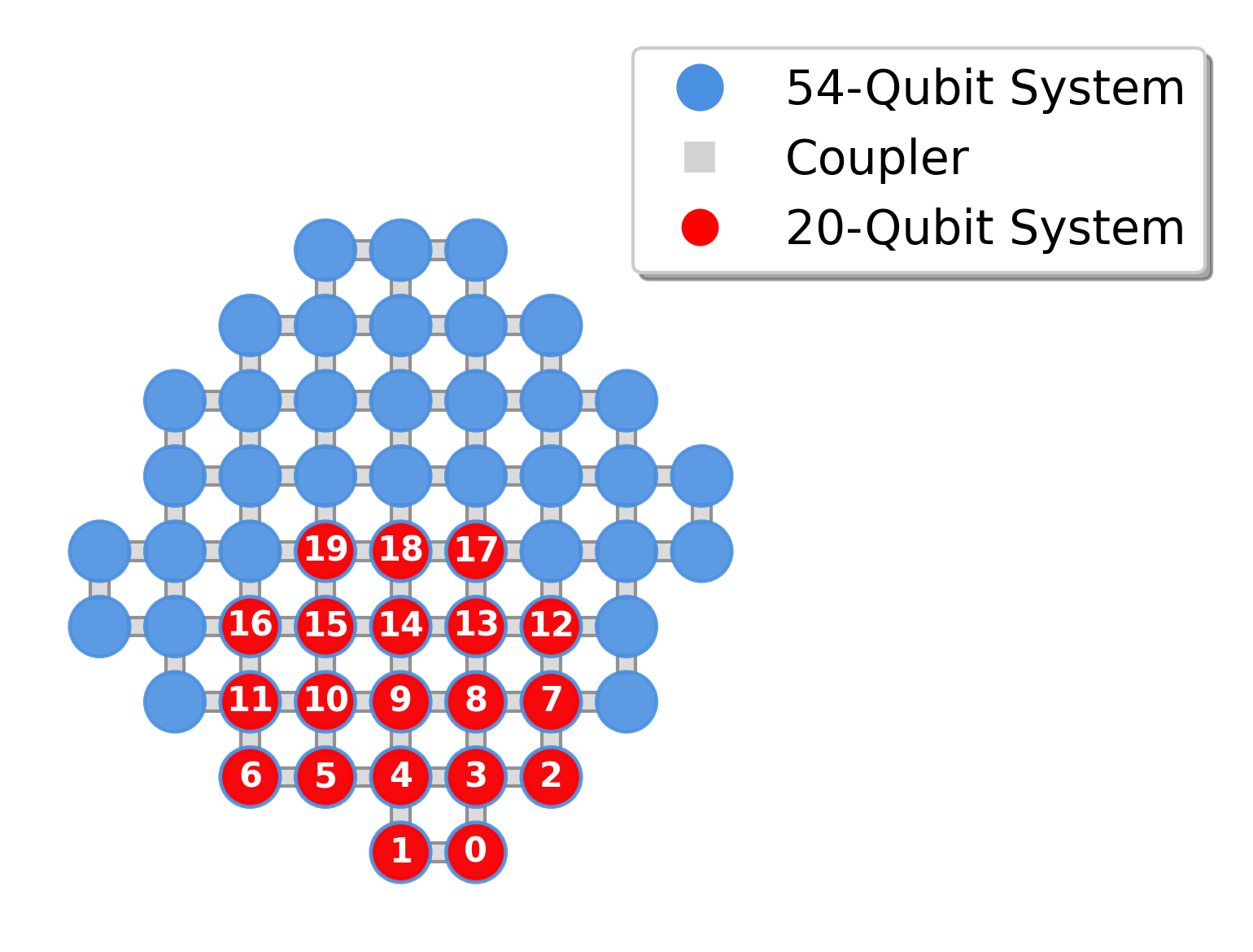}
        \caption{The layout of our in-house IQM 20-qubit QPU (red), also mapped onto the connectivity graph of the IQM 54-qubit QPU (blue). Between qubits are tunable couplers (gray).}
\label{fig:qpu_layout}
\end{figure}

Building on these efforts our contribution is a comprehensive, data-driven analysis of long-term calibration dynamics in an IQM 20-qubit superconducting quantum processor (Fig. \ref{fig:qpu_layout}) deployed in our HPC center\cite{10528924}, based on 250 days of daily calibrations. We systematically quantify temporal and cross-metric correlations, and apply unsupervised learning (including graph-based clustering) to classify qubits by stability and noise characteristics. We directly validate our findings with real hardware experiments using GHZ and Hadamard circuits. Our methods support practical circuit-to-hardware mapping and guide intelligent quantum device management in HPC centers.

The structure of the paper is as follows. Section~\ref{sec:calibration_data_basics} presents an empirical analysis of the dataset of calibration quality metrics. Section~\ref{sec:corr} compares multiple correlation and clustering techniques for device health monitoring. Section~\ref{sec:use} validates the analysis with experimental benchmarks. Section~\ref{sec:conclusion} concludes and outlines future directions.

\section{Calibration Quality Metrics}\label{sec:calibration_data_basics}
Our in-house quantum computer is hosted within our HPC computing cube and connected to SuperMUC-NG. It is a commercial system from IQM\cite{IQM_Garnet_QPU}. The quantum processing unit (QPU), dubbed {\it Q-Exa}, comprises 20 superconducting qubits arranged in a 2D grid, supporting nearest-neighbor coupling for high-fidelity two-qubit operations at zero temperature ($\sim 7 - 10$ millikelvin) in a dilution refrigerator. Fig.~\ref{fig:qpu_layout} shows our 20-qubit layout (red) as a contiguous subset of the IQM 54-qubit chip (blue). Each physical qubit is controlled via custom microwave and flux lines, enabling both single- and two-qubit gates. Calibration cycles produce key quality metrics for every qubit and pair, including relaxation time $T_1$, dephasing time $T_2$, readout fidelity and gate fidelities\cite{10.1063/1.5089550}. These metrics are extracted from standard quantum characterization protocols, with parameters obtained via nonlinear least square fits to physical models.

Over 250 days of operation, we collected  $250\times 20$ readings for each single-qubit metric and $250\times 30$ for two-qubit CZ fidelities. By maintaining a fixed calibration schedule (once per day) and consistent pulse/readout configurations, we ensure that metric fluctuations reflect true device and environmental variability, without procedural drift.

\subsection{\(T_1\) and \(T_2\)}
\begin{figure}[ht]
    \centering
    \begin{minipage}[t]{0.48\columnwidth}
        \centering
        \stackinset{l}{0.1em}{t}{0.1em}{\small\textbf{(a)}}{
            \includegraphics[width=1\linewidth]{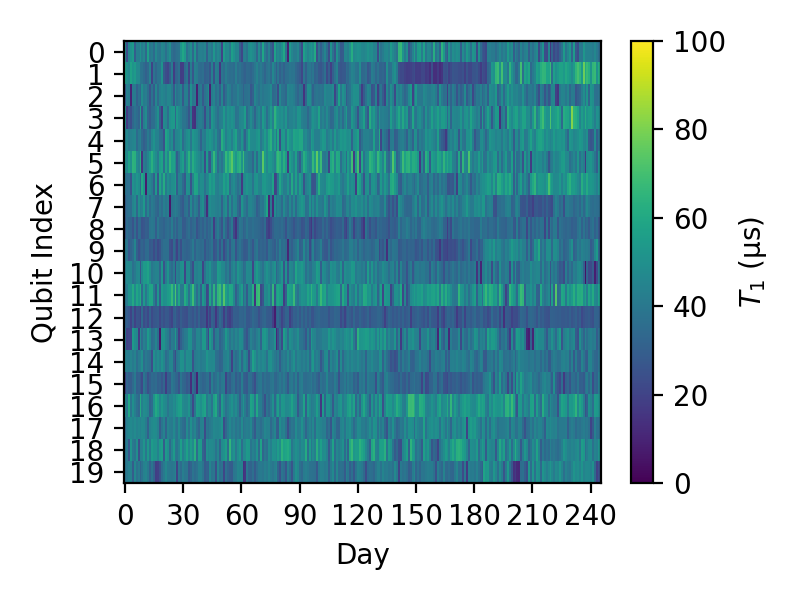}
        }
        \label{fig::t1_heatmap}
    \end{minipage}
    \hfill 
    \begin{minipage}[t]{0.48\columnwidth}
        \centering
        \stackinset{l}{0.1em}{t}{0.1em}{\small\textbf{(b)}}{
            \includegraphics[width=1\linewidth]{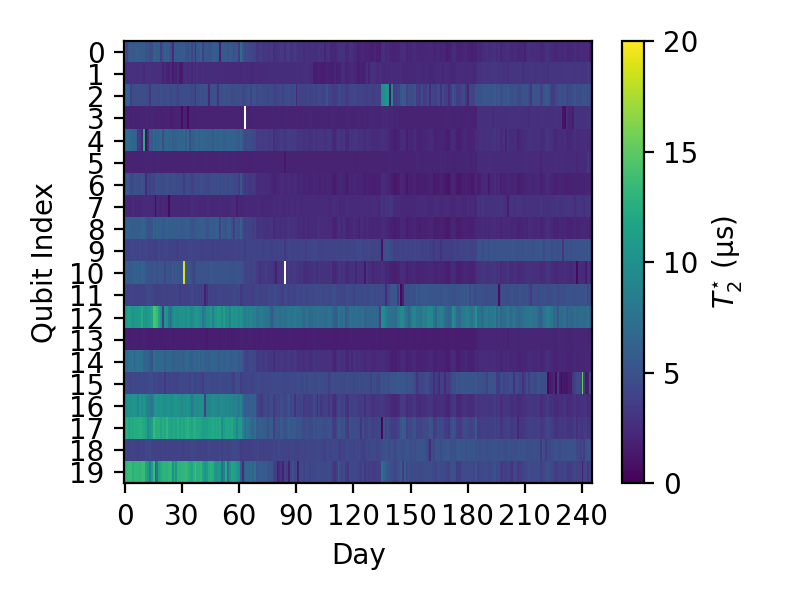}
        }
        \label{fig::t2_heatmap}
    \end{minipage}
\hfill
    \begin{minipage}[t]{0.48\columnwidth}
        \centering
        \stackinset{l}{0.1em}{t}{0.1em}{\small\textbf{(c)}}{
            \includegraphics[width=1\linewidth]{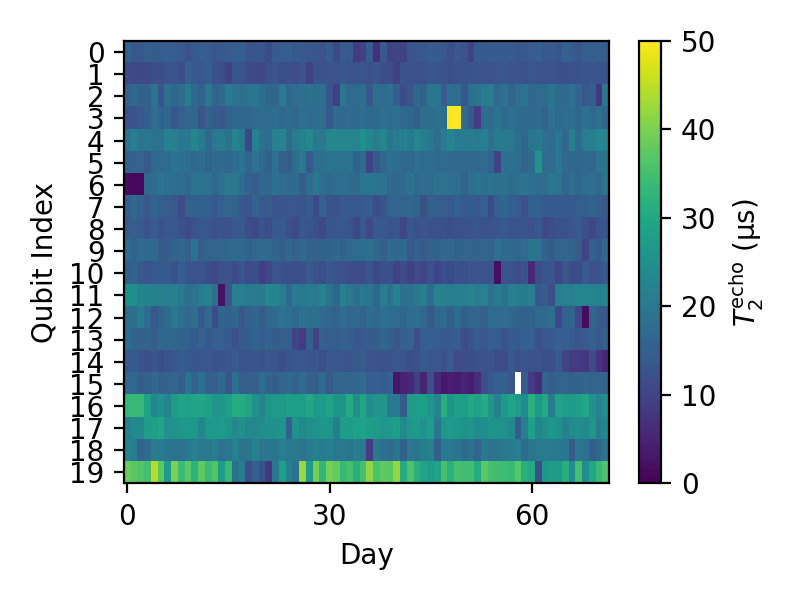}
	}
        \label{fig::t1_std}
    \end{minipage}
    \hfill 
    \begin{minipage}[t]{0.48\columnwidth}
        \centering
        \stackinset{l}{0.1em}{t}{0.1em}{\small\textbf{(d)}}{
            \includegraphics[width=1\linewidth]{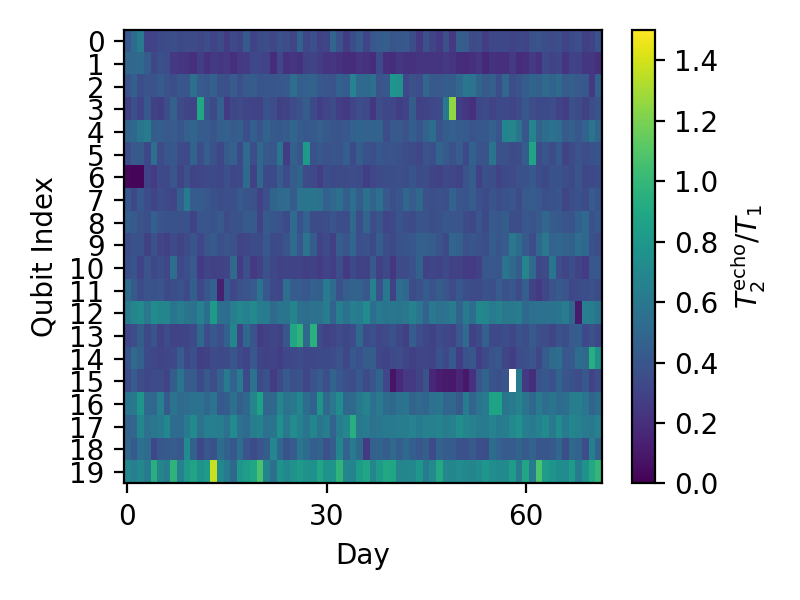}
        }
        \label{fig::t2star_std}
    \end{minipage}

    \caption{(a) Relaxation time $T_1$ heatmap over time. The average across 20 qubits and over more than $250$ days shows $40.95 \pm 10.54 (\textrm{std}) \mu$s. (b) Dephasing time (Ramsey) $T^{\star}_2$ heatmap over time, with the average $3.89\pm 2.27\mu$s. (c) Spin echo $T^{\rm echo}_2$ heatmap over time, with the average $17.7\pm 6.05\mu$s. (d) $T^{\rm echo}_2/T_1$ heatmap over time. The mean is around $0.44$ with std $0.16$. The minimal ratio is $0.03$ and the maximal is $1.4$. The ratios larger than 2.0 are omitted. 
    Color scales are $(0,100)\mu$s, $(0,20)\mu$s, $(0,50)\mu$s and $(0,1.5)$ (dimensionless) for $T_1$, $T^{\star}_2$, $T^{\rm echo}_2$ and $T_2^{\rm echo}/T_1$, respectively. In our data set $T^{\rm echo}_2$ was recorded half a year later than $T^{\star}_2$.}
    \label{fig:combined_t1_t2}
\end{figure}

The relaxation time $T_1$ is determined by the excitation decay experiment, and extracted from exponential decay fits to excited state population measurements, $P_1(t)=A\exp{(-t/T_1)}+B$\cite{10.1063/1.5089550}. Dephasing times $T_2$ are measuring phase decoherence time of superposition state and using Ramsey ($T^{\star}_2$) and spin echo ($T^{\rm echo}_2$) protocols, fitting $P_2(t)=A\exp{(-t/T_2)}\cos{(\omega t+\phi)}+B$\cite{10.1063/1.5089550}. Fig.~\ref{fig:combined_t1_t2} shows heatmaps for $T_1$, $T_2^{\star}$, $T_2^{\rm echo}$, and the ratio $T_2^{\rm echo}/T_1$ over 250 days.
These heatmaps present a overview of both temporal evolution and spatial heterogeneity across the full 20-qubit device.

We monitor $T_1$ and $T_2$ for sudden drops, persistent decreases, or large standard deviations (std), that are indicators of instability due to environmental drift or microscopic defects, e.g., TLS. A qubit with specific frequency may interact with TLS in resonance with time.
Qubits with large standard deviations in $T_1$ and $T_2$ are flagged as potentially unstable. In our data, the mean values over 250 days are typically $T_1\approx 41\mu$s, $T^{\star}_2\approx4\mu$s, and $T^{\rm echo}_2\approx18\mu$s. $T_2^{\star}$ is sensitive to both slow noise sources and static disorder, and is often much shorter than $T_1$. The spin echo dephasing time $T_2^{\rm echo}$ offers an improved measure of phase coherence by refocusing low-frequency noise and then canceling it. Here the color map shows that in echo experiments coherence for some qubits is significantly extended.

The ratio $T^{\rm echo}_2/T_1$ is a sensitive indicator of diagnosing the balance between energy relaxation and dephasing. The energy relaxation time $T_1$ is sensitive to energy dissipation. The decoherence time $T_2$ is limited by both energy relaxation and pure dephasing $T^{-1}_2=\frac{1}{2}T^{-1}_1 + T^{-1}_{\phi}$, where $T_{\phi}$ is the pure dephasing time involving phase randomization, and $T_2$ here refers to $T^{\rm echo}_2$. Thus, the ratio $T_2/T_1\leq 2$. In ideal case, if pure dephasing is negligible, $T_2/T_1\approx 2$. Decoherence is mostly due to energy relaxation, and the qubit's coherence is limited by $T_1$ only. In real systems, additional dephasing mechanisms also play a role, so $T_2/T_1<2$. For our device the mean ratio $T_2/T_1\approx 0.44$. Short $T_2$s suggest the pure dephasing is significant, that is likely caused by $1/f$ noise, flux fluctuations, TLS, or control imperfections. Fig.~\ref{fig:combined_t1_t2}(d) depicts this ratio, showing spatial and temporal variability across the QPU.

\subsection{Readout Fidelity}
Readout fidelity measures the probability of correctly identifying the computational basis state $\lvert0\rangle$ and $\lvert1\rangle$ during measurement. It is determined as $
F_{\rm readout} \;=\;\frac{P(0\,|\,0)+P(1\,|\,1)}{2}\;=\;1-\frac{P(0\,|\,1)+P(1\,|\,0)}{2},
$
where \(P(b\,|\,a)\) is the probability of measuring outcome \(b\) given that the qubit was prepared in state \(\lvert a\rangle\). The metric is extracted directly from repeated preparations and measurements, without the need for curve fitting.

\begin{figure}[ht]
    \centering
    \begin{minipage}[t]{0.48\columnwidth}
        \centering
        \stackinset{l}{0.1em}{t}{0.1em}{\small\textbf{(a)}}{
            \includegraphics[width=1\linewidth]{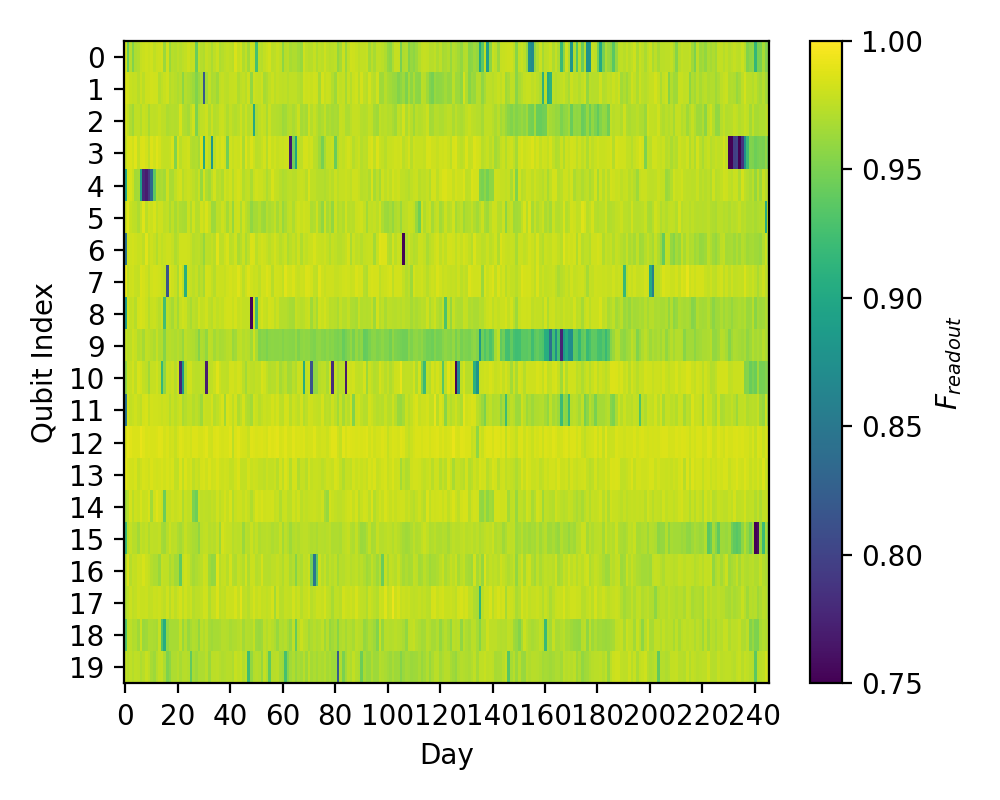}
        }
        \label{fig:readout_heatmap}
    \end{minipage}
    \hfill 
    \begin{minipage}[t]{0.48\columnwidth}
        \centering
        \stackinset{l}{0.1em}{t}{0.1em}{\small\textbf{(b)}}{
            \includegraphics[width=1\linewidth]{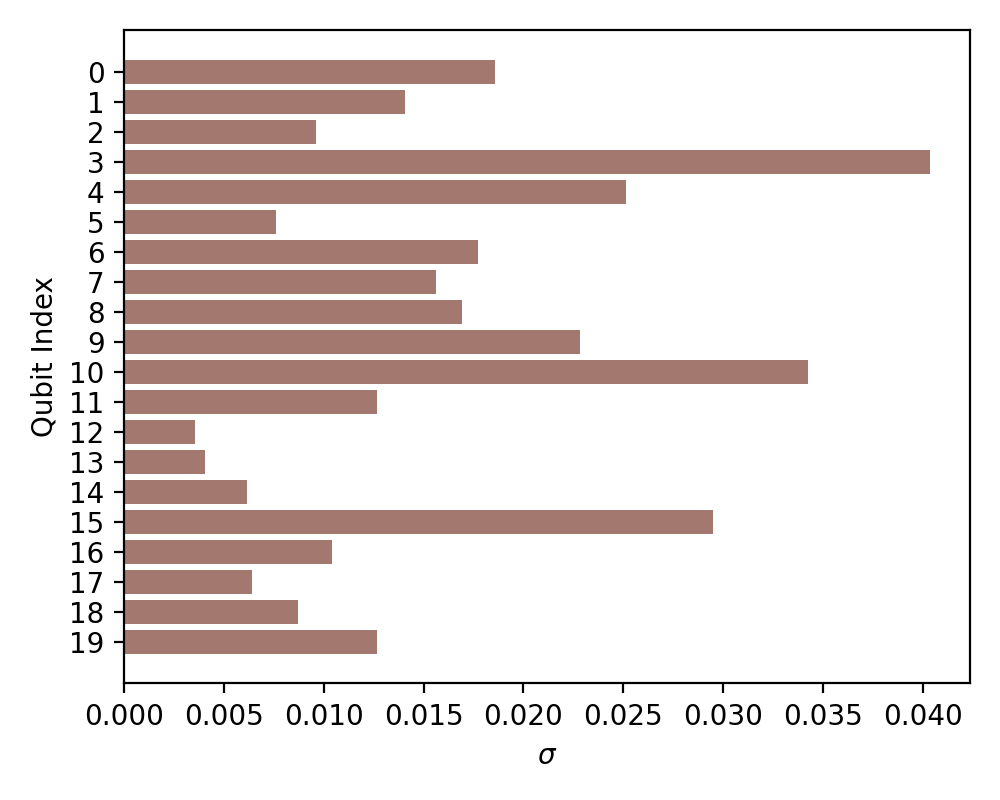}
        }
        \label{fig:readout_std_bar}
    \end{minipage}
    \caption{(a) Heatmap of readout fidelity for each qubit across 250 days.  Color scale ranges from 0.75 to 1.00.  The overall mean fidelity is \({0.972}{\pm 0.020}\). (b) Bar-plot of standard deviation $\sigma$ along $250$ days for each qubit. We can see which qubits are most unstable in readout fidelity.}
    \label{fig:combined_readout}
\end{figure}

Figure~\ref{fig:combined_readout} (a) presents a heatmap of daily readout fidelities for each qubit 20 across 250 calibration cycles. The color scale $0.75-1.00$ shows both temporal drift and device heterogeneity. Most qubits maintain high-fidelity performance $F_{readout}\approx0.972\pm0.020$, but certain qubits, e.g., qubit $9$ displays reduced fidelity. Figure~\ref{fig:combined_readout}(b) summarizes the standard deviation of readout fidelity per qubit, clearly identifying those with persistent instability. Sudden drops in the heatmap can signal intermittent hardware issues (e.g., amplifier drift, spurious interference), while long-term trends may reflect slow degradation or environmental change. Clusters of qubits with similar patterns with big variance suggest non-uniformities in the readout chain. This motivates the use of unsupervised clustering methods in subsequent sections, alongside other calibration metrics.

\subsection{Single- and Two-Qubit Gate Fidelity}
Gate fidelity measures the accuracy of an implemented quantum gate relative to its ideal unitary. For single-qubit gates ($F_{1q}$) we use randomized benchmarking (RB) to fit the decay of the survival probability over $m$ Clifford operations, $P(|0\rangle,m)=Ap^m+B$. The average gate fidelity is then $F_{1q}=(1+p)/2$. For two-qubit CZ gates, interleaved RB is used to fit $P(|00\rangle,m)=Ap^m+B$, yielding $F_{2q}=(1+3p)/4$. Once $p$ is extracted from the exponential decay of RB fittings, the fidelity can be obtained\cite{10.1063/1.5089550}.

\begin{figure}[ht]
\centering
    \begin{minipage}[t]{0.48\columnwidth}
        \centering
        \stackinset{l}{0.1em}{t}{0.1em}{\small\textbf{(a)}}{
            \includegraphics[width=1\linewidth]{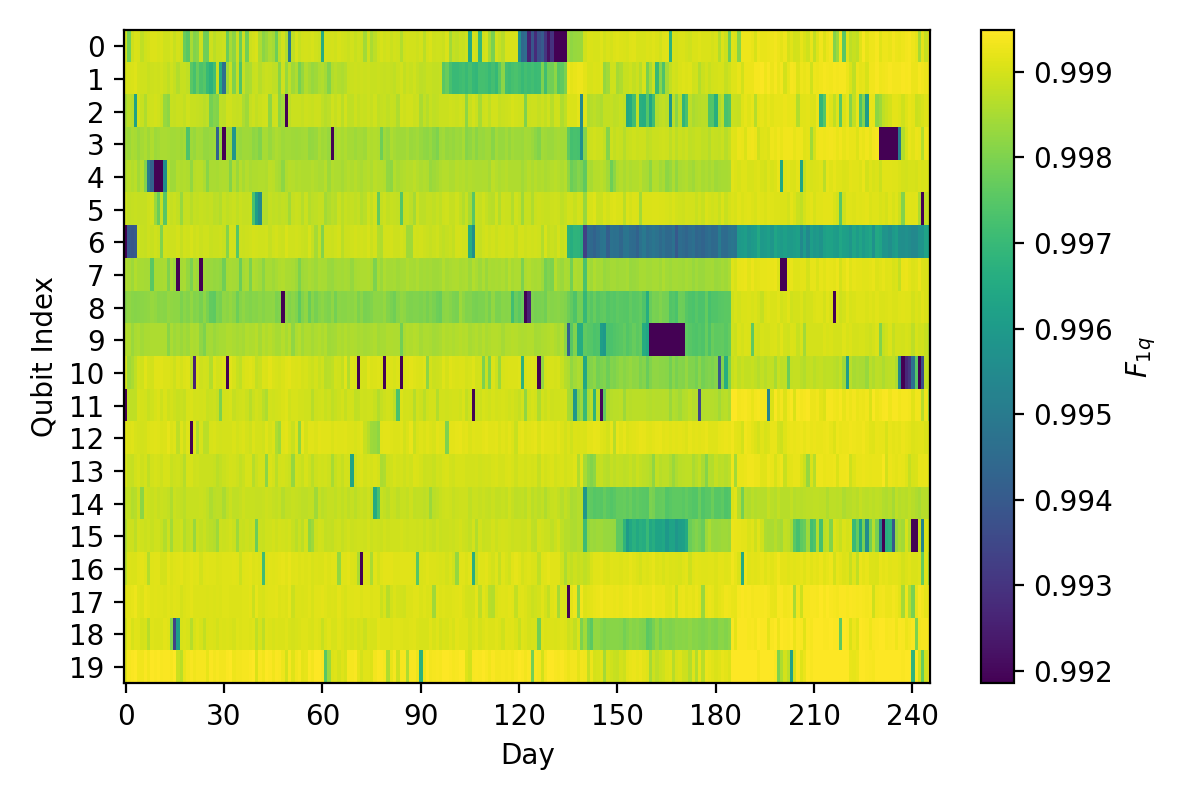}
        }
        \label{fig:oneq_heatmap}
    \end{minipage}
    \hfill 
    \begin{minipage}[t]{0.48\columnwidth}
        \centering
        \stackinset{l}{0.1em}{t}{0.1em}{\small\textbf{(b)}}{
            \includegraphics[width=1\linewidth]{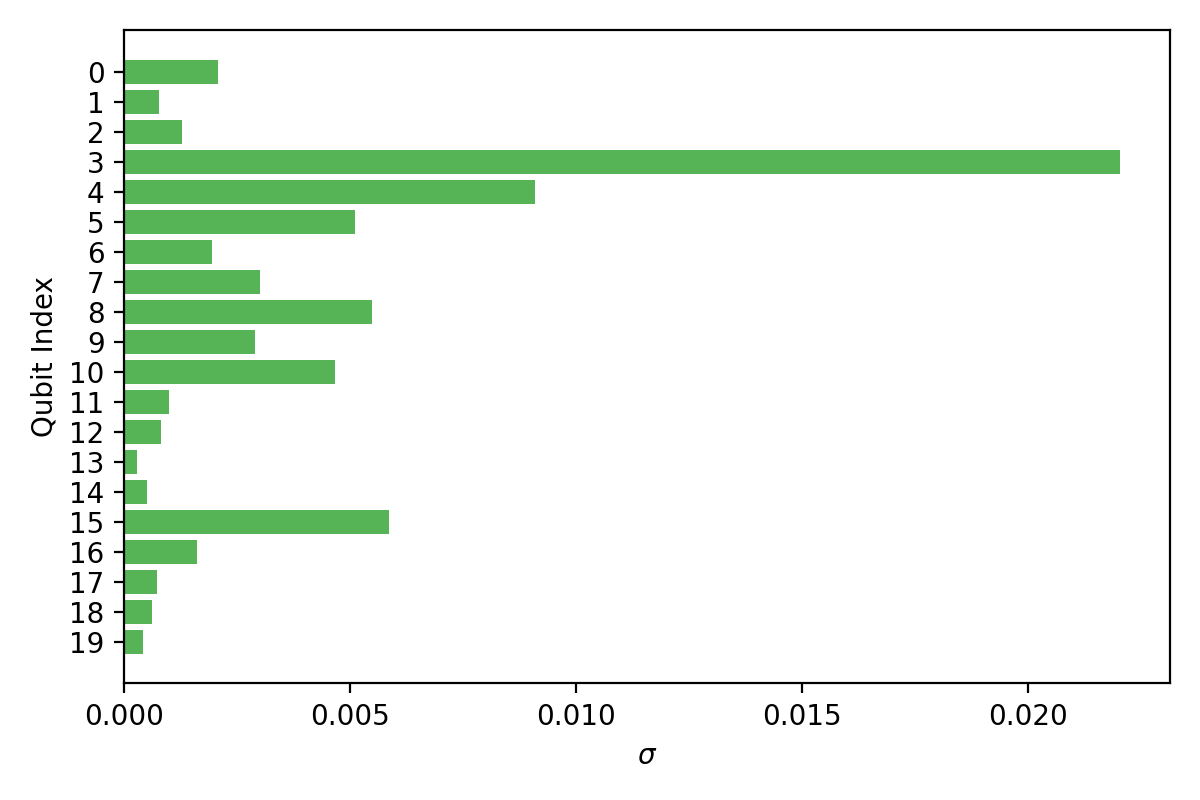}
        }
        \label{fig:oneq_std}
    \end{minipage}
\hfill
    \begin{minipage}[t]{0.48\columnwidth}
        \centering
        \stackinset{l}{0.1em}{t}{0.1em}{\small\textbf{(c)}}{
            \includegraphics[width=1\linewidth]{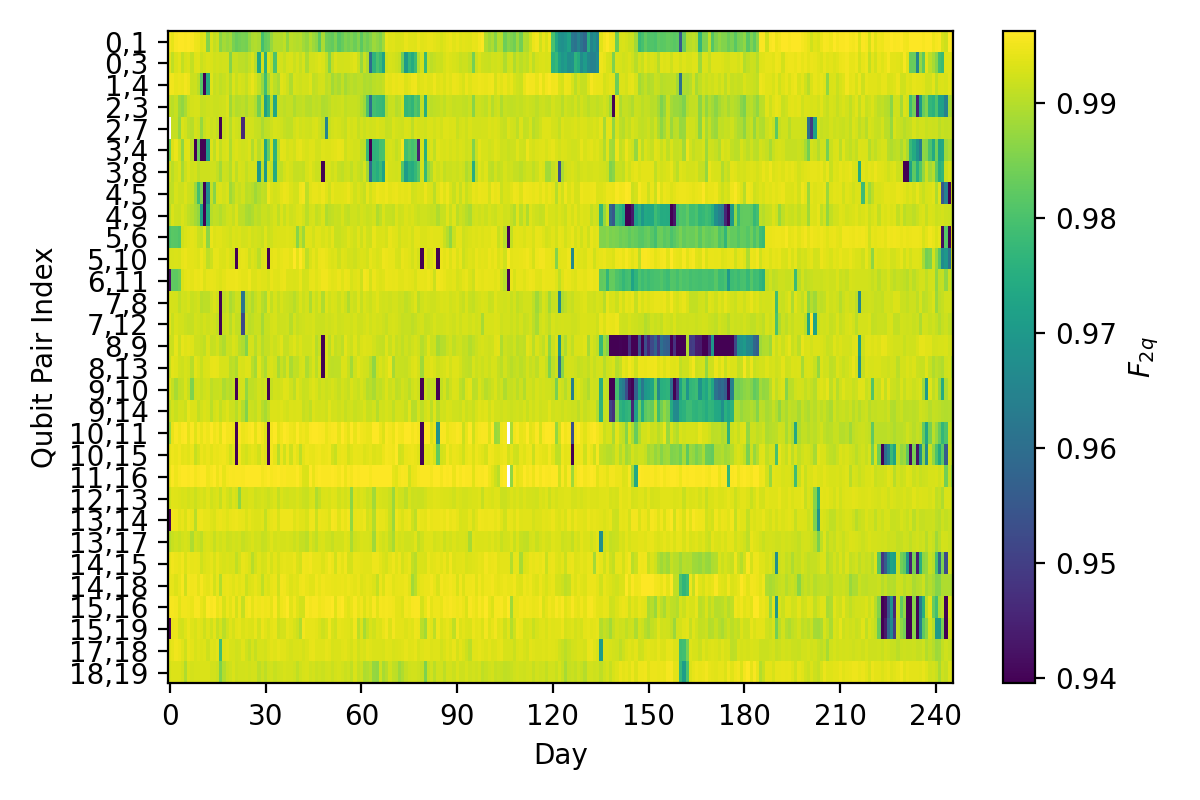}
	}
        \label{fig:twoq_heatmap}
    \end{minipage}
    \hfill 
    \begin{minipage}[t]{0.48\columnwidth}
        \centering
        \stackinset{l}{0.1em}{t}{0.1em}{\small\textbf{(d)}}{
            \includegraphics[width=1\linewidth]{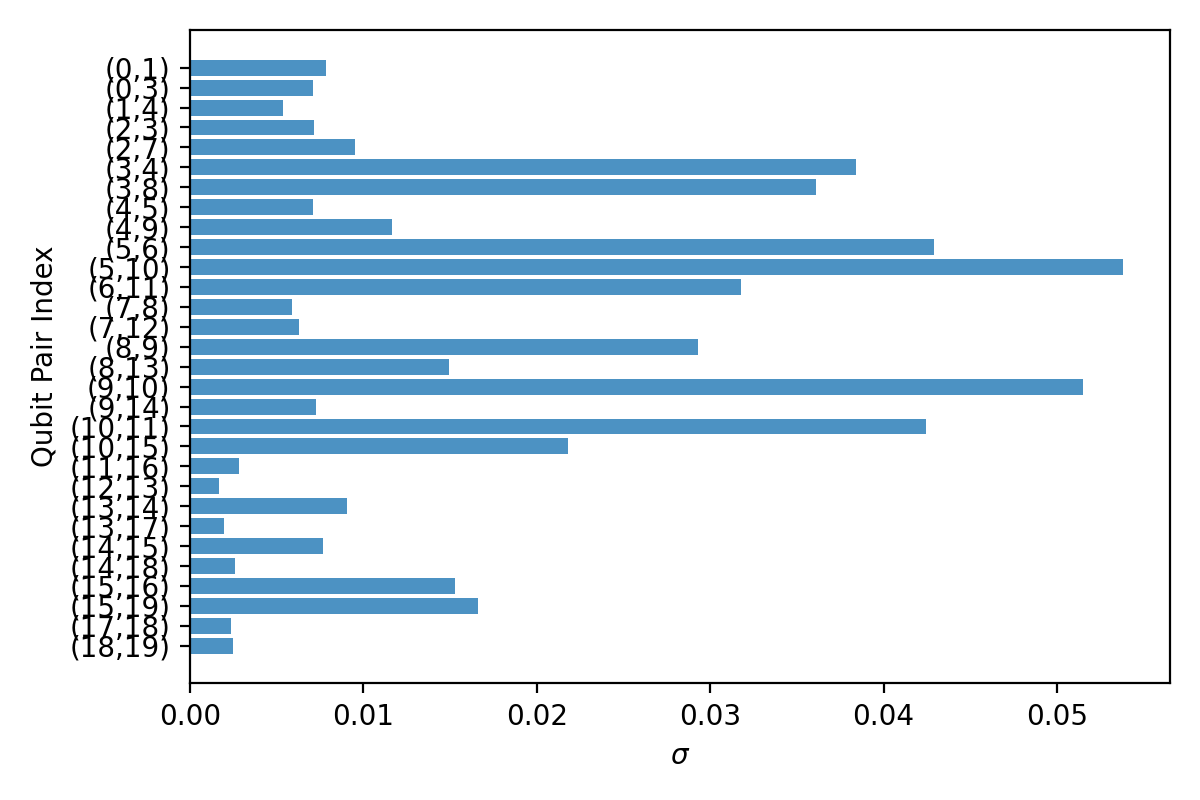}
        }
        \label{fig::twoq_std}
    \end{minipage}
    \caption{Heatmaps of (a) single‐qubit and (c) two-qubit CZ gate fidelities over 250 days and their bar-plots of variance over time for single-qubit gate (b) and two-qubit gate (d).  From (a) and (c) we see clearly two boundaries that correspond to two warm-ups during the whole time. Single-qubit gate of qubit 3 shows a high variability. The two-qubit gates related to qubits $3, 5, 10$ have high variabilities. Color scales run from $0.99$ to $1$ for single‐qubit, and $0.95$ to $1$ for two-qubit fidelities.}
    \label{fig:gate_heatmap}
\end{figure}

Figure~\ref{fig:gate_heatmap} summarizes daily gate fidelities across the device. Subfigures (a) and (c) present heatmaps for all single qubits and two-qubit pairs, while (b) and (d) show the per-qubit/pair standard deviation as bar plots. Most single-qubit gates maintain very high fidelity $F_{1q}=0.9983\pm0.0035$ (std), with rare drops possibly due to environmental events or control signal retuning. However, Figure~\ref{fig:gate_heatmap} (a) shows that qubit 6 has a sharpest drop in gate fidelity while qubit 3 exhibits persistent variability. Two-qubit CZ fidelities are lower, $F_{2q}=0.9895\pm0.0167$ (std), and exhibit larger fluctuations over time. Notably, pairs involving qubits 3, 5, 10 are consistently less stable, with standard deviation up to $\sim 0.05$. This may be related to local crosstalk, flux noise, or transient coupling instabilities. 

Both heatmaps (a) and (c) show two pronounced boundaries in time (at about Day 130 and Day 180), corresponding to two times cryogenic warm-up and cool-down circles for our chip maintenance. Such global events temporarily degrade gate performance  across large area of the chip. The bar-plots in (b) and (d) directly show the noisiest qubits and couplers, pointing to the most unstable regions for further diagnostic attention. Spatial and temporal patterns in gate fidelity as shown by these heatmaps and variances actually build a dynamic performance envelope for quantum circuits submitted on the chip. 

\subsection{Probability Distribution of Calibration Metrics}
\begin{figure}[ht]
    \centering
    \begin{minipage}[t]{0.48\columnwidth}
        \centering
        \stackinset{l}{0.1em}{t}{0.1em}{\small\textbf{(a)}}{
            \includegraphics[width=1\linewidth]{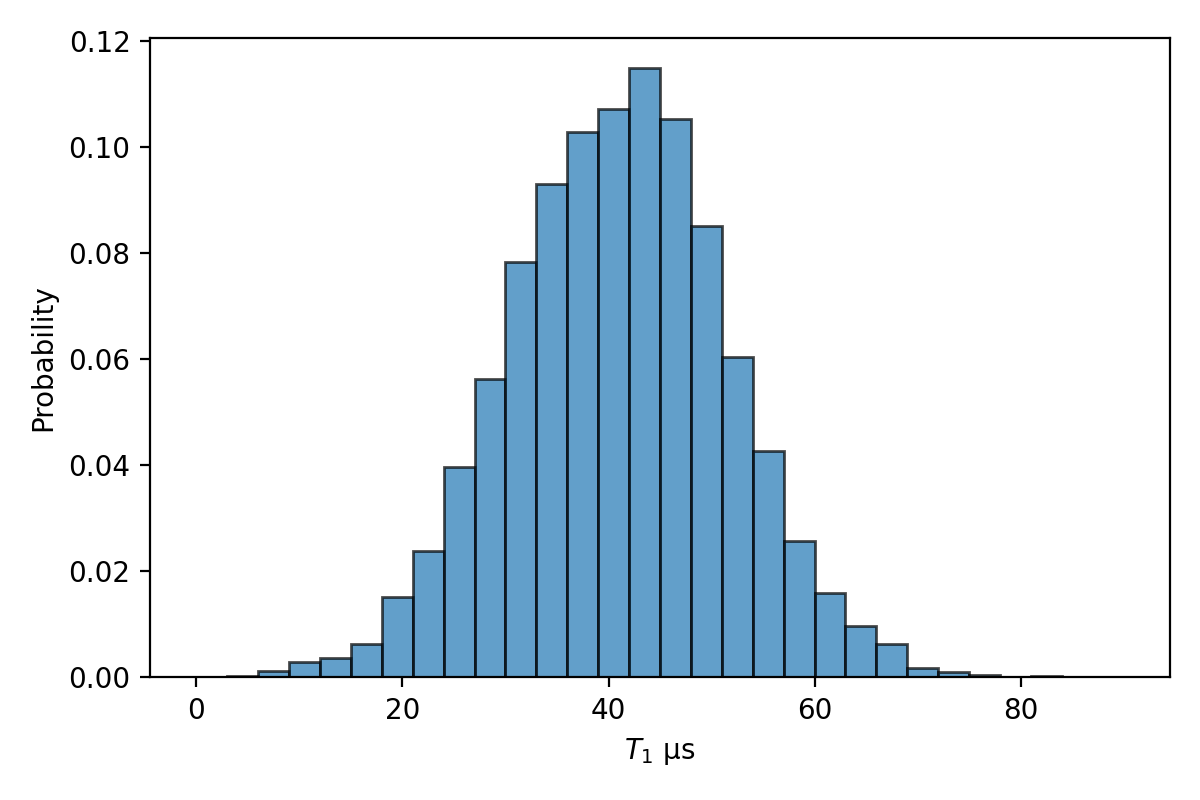}
	}
        \label{fig:dist_t1}
    \end{minipage}
    \hfill 
    \begin{minipage}[t]{0.48\columnwidth}
        \centering
        \stackinset{l}{0.1em}{t}{0.1em}{\small\textbf{(b)}}{
            \includegraphics[width=1\linewidth]{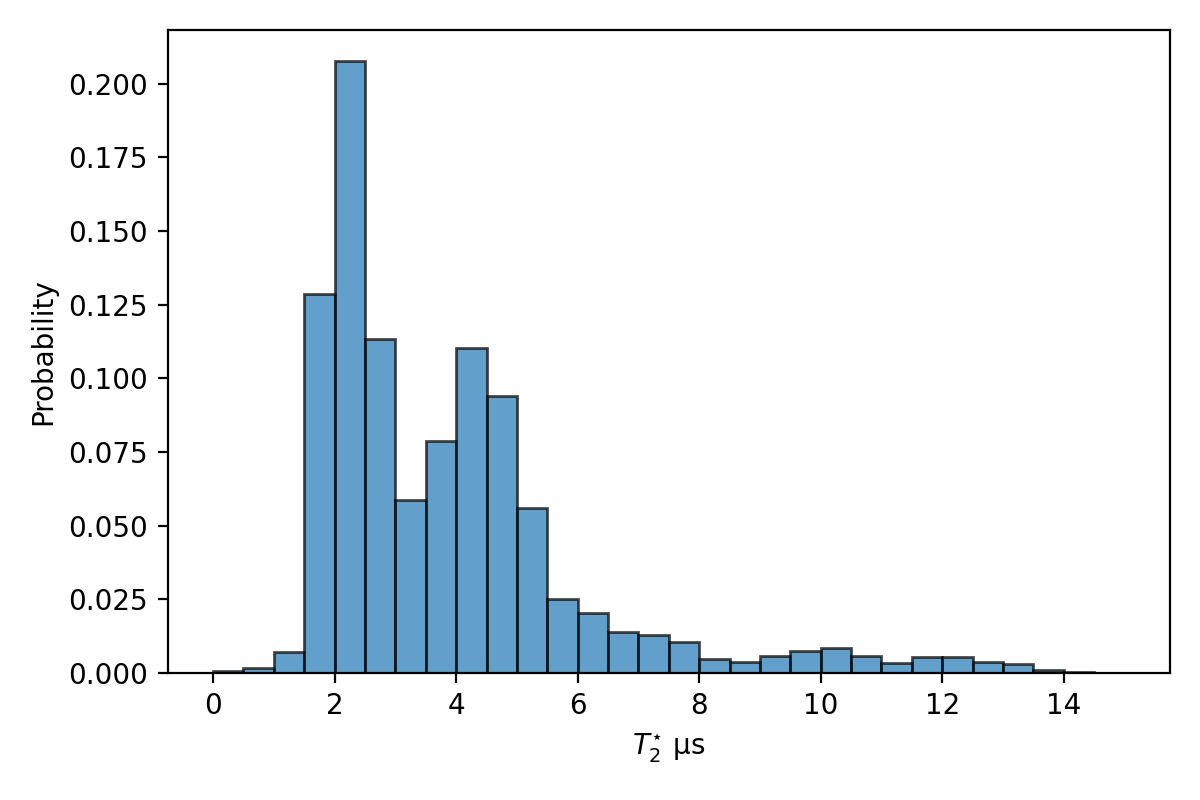}
        }
        \label{fig::dist_t2}
    \end{minipage}
\hfill
    \begin{minipage}[t]{0.48\columnwidth}
        \centering
        \stackinset{l}{0.1em}{t}{0.1em}{\small\textbf{(c)}}{
            \includegraphics[width=1\linewidth]{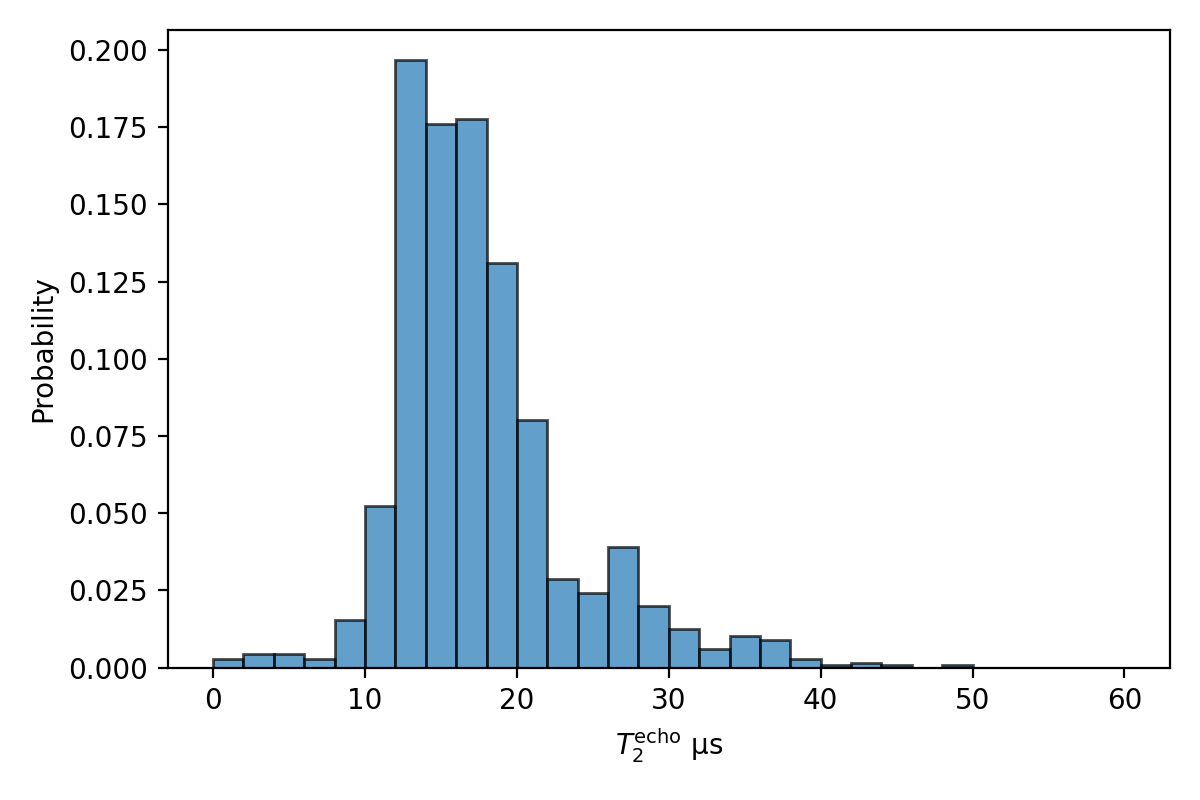}
	}
        \label{fig:dist_t2echo}
    \end{minipage}
    \hfill 
    \begin{minipage}[t]{0.48\columnwidth}
        \centering
        \stackinset{l}{0.1em}{t}{0.1em}{\small\textbf{(d)}}{
            \includegraphics[width=1\linewidth]{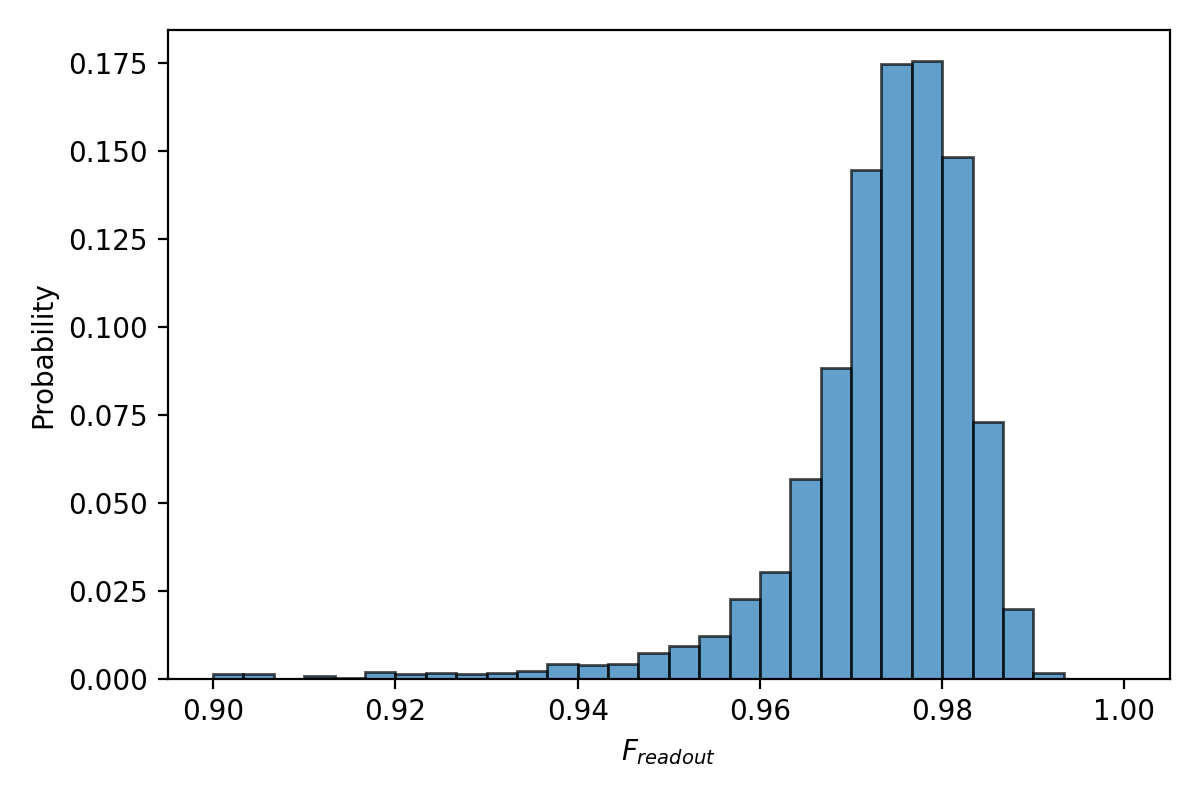}
        }
        \label{fig::dist_readout}
    \end{minipage}
\hfill 
    \begin{minipage}[t]{0.48\columnwidth}
        \centering
        \stackinset{l}{0.1em}{t}{0.1em}{\small\textbf{(e)}}{
            \includegraphics[width=1\linewidth]{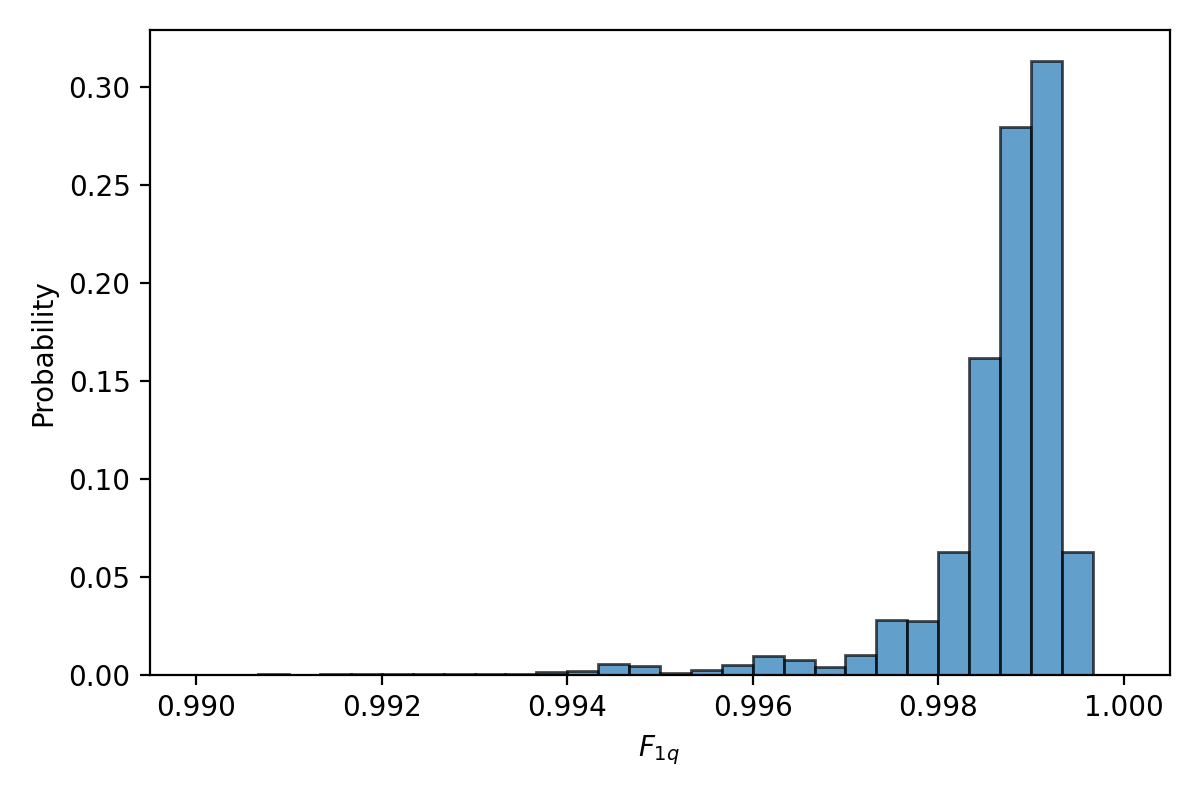}
        }
        \label{fig::dist_oneq}
    \end{minipage}
\hfill 
    \begin{minipage}[t]{0.48\columnwidth}
        \centering
        \stackinset{l}{0.1em}{t}{0.1em}{\small\textbf{(f)}}{
            \includegraphics[width=1\linewidth]{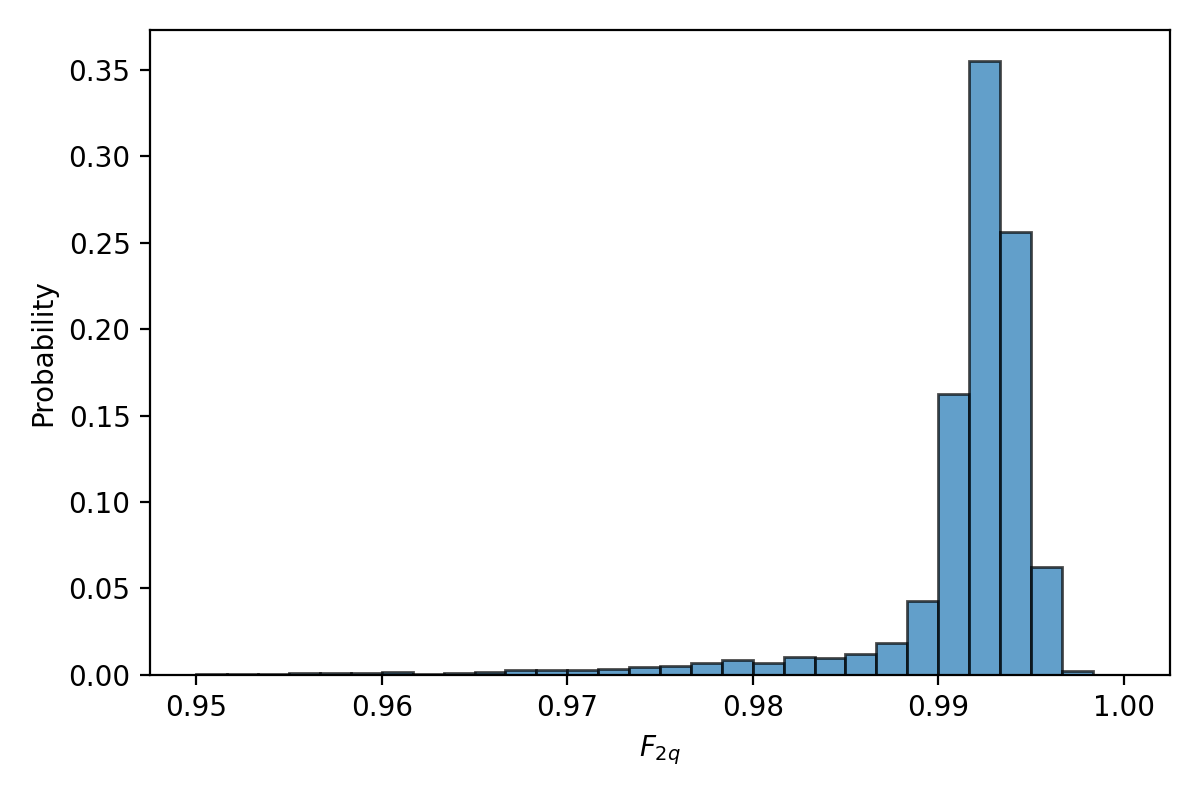}
        }
        \label{fig::dist_twoq}
    \end{minipage}

    \caption{Probability distribution of each quality metric across all qubits/pairs and days: (a) $T_1$, (b) $T^{\star}_2$, (c) $T^{\rm echo}_2$, (d) $F_{readout}$, (e) $F_{1q}$, and (f) $F_{2q}$. $T_1$ is relatively stable, while all other other metrics show fat-tailed or even bi-modal ($T^{\star}_2$) distributions.}
    \label{fig:probability_pmf}
\end{figure}

Beyond time evolution, understanding the underlying statistical distributions of key quality metrics is crucial for robust device monitoring and anomaly detection. Fig.~\ref{fig:probability_pmf} shows the empirical (binned) probability density functions for $T_1$, $T_2$, $T^{\rm echo}_2$, readout fidelity, and both single- and two-qubit gate fidelities, summarized over all qubits and calibration cycles in a representative time window. $T_1$ exhibits a near-Gaussian profile, centered at $\sim 40\mu$s with relatively thin tails. This is expected. $T_1$ is dominated by many small, uncorrelated decay process which yield near-normal variability by the central limit theorem. $T^{\star}_2$ and $T^{\rm echo}_2$ distributions are notably right-skewed, with long tails extending to higher values. $T^{\star}_2$ hints at bi-modal shapes. Occasional high $T_2$ events occur when low-frequency noise is temporarily suppressed, while the bulk of the distribution remain below the $T_1$ limit. Readout and gate fidelities cluster very tightly near $1.0$, but both exhibit heavy lower tails. These outlier events often signal transient defects, such as TLS resonance, readout amplifier drift, or crosstalk, that degrade performance on some day or specific channels. The distortion of control pulses may cause these heavy lower tails. Individual TLS defects can randomly hop in and out resonance. If an individual TLS defect couples strongly to certain qubit frequencies, it will degrade their coherence and gate performance for days until the TLS spectrally hops out.

The existence of heavy lower tails in fidelity and coherence distributions emphasizes the need for robust anomaly detection and automated health monitoring routines. By modeling these tail probabilities one can set risk-aware operational thresholds beyond the simple threshold settings. These distributions incorporated into qubit/pair feature vectors can further sharpen the clustering distinction.

\section{Correlations}\label{sec:corr}
In principle, a well-calibrated quantum processor should yield quality metrics that are statistically independent, reflecting only random noise. However, real-world systems rarely achieve this ideal target. Environmental factors (temperature, vibrations, magnetic fields, etc.), hardware drift, and persistent defects such as TLS can introduce temporal and cross-metric correlations, even under regular recalibration. Incremental recalibration schemes that rely on prior parameter values may further amplify these effects, causing subtle trends or dependencies that may accumulate over time. In this section, we systematically investigate both temporal autocorrelation and cross-metric correlation in our calibration data, and explore how unsupervised learning methods can reveal latent structure among qubits and their operational metrics. 

\subsection{Temporal Autocorrelation}
To quantify how much each calibration metric remembers its past, we analyze its temporal autocorrelation function (ACF) following the classic methodology\cite{box2015time, shumway2017time, vishwas2020hands}. A significant ACF indicates that today's calibration value can be predicted at least in part from its own history and there may be drift, periodic trends or other memory effects, whereas a rapidly vanishing ACF implies a memoryless or white-noise process. Such temporal dependencies are important for identifying slow environmental drift or latent hardware issues. For instance, $T_1$ and $T_2$ may show long-range correlations due to environmental changes. Gate fidelity ACF may mean pulse calibration drift. Readout ACFs can reflect resonator frequency detuning or amplifier instability. Detecting these effects is crucial for robust predictive modeling and anomaly detection.

\begin{figure}[ht]
    \centering
    \begin{minipage}[t]{0.48\columnwidth}
        \centering
        \stackinset{l}{0.1em}{t}{0.1em}{\small\textbf{(a)}}{
            \includegraphics[width=1\linewidth]{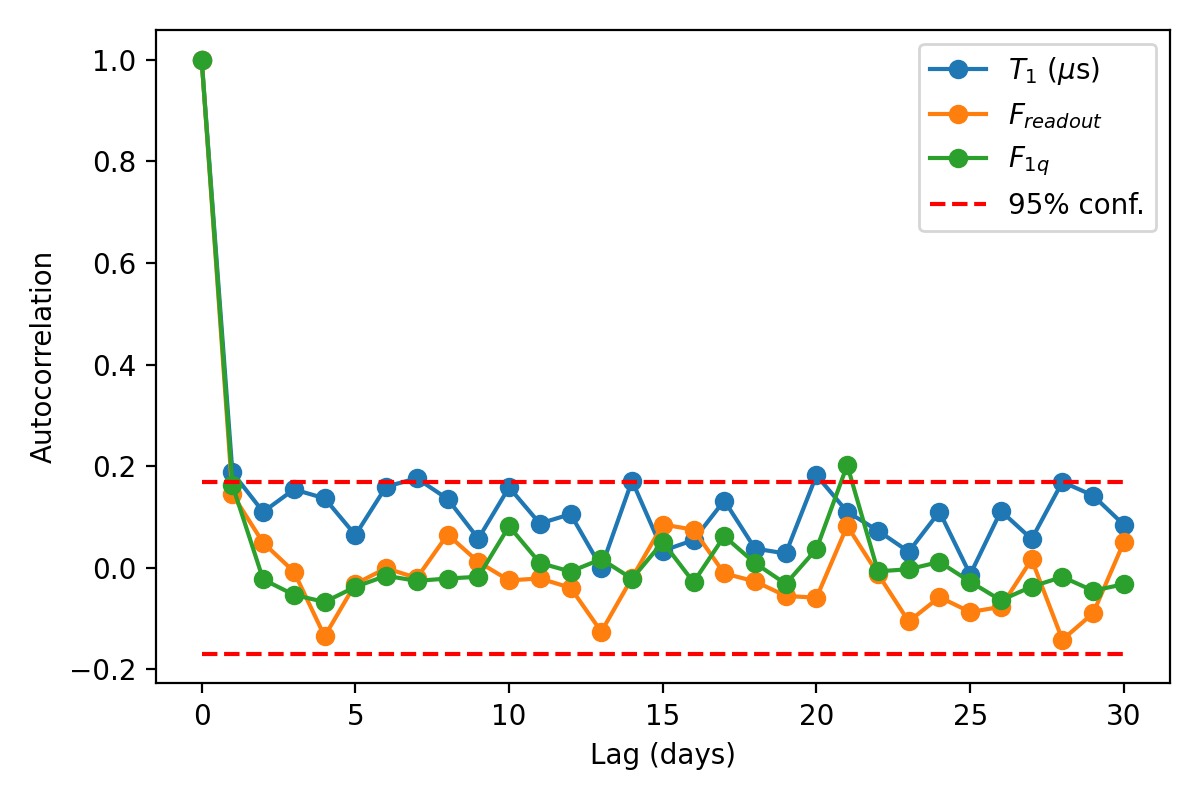}
	}
        \label{fig:acf_mean}
    \end{minipage}
    \hfill 
    \begin{minipage}[t]{0.48\columnwidth}
        \centering
        \stackinset{l}{0.1em}{t}{0.1em}{\small\textbf{(b)}}{
            \includegraphics[width=1\linewidth]{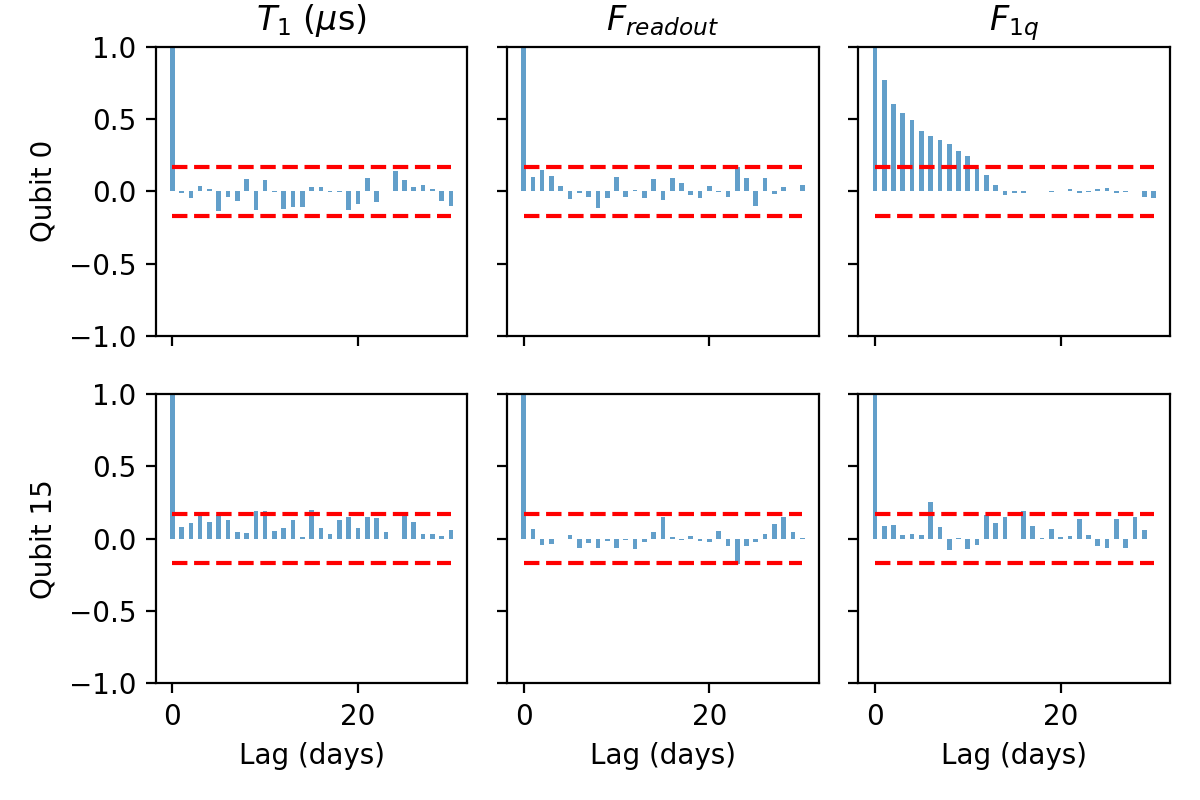}
        }
        \label{fig:acf_rep}
    \end{minipage}
 \caption{(a) ACF of daily average for metrics $T_1$, $F_{readout}$ and $F_{1q}$ across all 20 qubits with $95\%$ confidence interval. (b) Representative single-qubit ACFs for qubits $0$ and $15$. Qubit 0's single-qubit gate exhibits a long-lived drift for several days.}
    \label{fig:acf}
\end{figure}

For each metric, we calculate the ACF up to 30 days lag and $95\%$ confidence interval to distinguish statistically significant correlations. 
Fig.~\ref{fig:acf}(a) shows the ACF for daily means of $T_1$, $F_{\mathrm{readout}}$, and $F_{1q}$ across all qubits. Notably, readout and single-qubit gate fidelities lose significant memory after just one day, whereas $T_1$ shows moderate multi-day persistence, likely reflecting environmental coupling or underlying hardware drift. Fig.~\ref{fig:acf}(b) illustrates representative single-qubit ACFs, showing diverse behavior across the chip. For instance, qubit 0’s single-qubit gate fidelity exhibits long-lived correlations, while qubit 15 appears nearly memoryless.

As a complementary summary, Table~\ref{tab:acf_stats} reports the mean and standard deviation of ACFs across the device at lags 1, 7, and 14 days for each metric, giving a device-wide view of typical memory length and variability. These empirical findings give operational suggestions: metrics with short memory support rapid recalibration, while those with longer timescales may require more adaptive maintenance or calibration schedules.

\begin{table}[ht]
  \centering
  \begin{tabular}{lccc}
    \toprule
    \textbf{Metric} & \textbf{ACF(1)} & \textbf{ACF(7)} & \textbf{ACF(14)} \\
    \midrule
    $T_{1}$               & \(0.10 \pm 0.10\) & \(0.02 \pm 0.09\) & \(0.01 \pm 0.08\) \\
    $F_{\rm readout}$     & \(0.16 \pm 0.23\) & \(0.07 \pm 0.15\) & \(0.04 \pm 0.12\) \\
    $F_{1q}$              & \(0.20 \pm 0.25\) & \(0.06 \pm 0.16\) & \(0.01 \pm 0.07\) \\
    \bottomrule
  \end{tabular}
  \caption{Mean and standard deviation of the autocorrelation function at lags 1, 7, and 14 days for each calibration metric.}
  \label{tab:acf_stats}
\end{table}

\subsection{Cross-Metric Correlation}
Understanding how different calibration metrics are correlated with each other is important for both diagnosis and maintenance decision-making as well. While each correlation method has its own strengths and quantum calibration data is typically nonlinear, non-Gaussian and exhibits complex dependencies, we thus systematically compare four methods to capture a detailed picture. The four methods we are using are Pearson, Spearman, distance correlation, and mutual information\cite{hastie2009elements}. Pearson is for linear relationship and assumes a normal distribution. Spearman is for monotonic trends, including linear and nonlinear. Distance correlation can detect arbitrary dependence, and the correlation is zero if and only if the two variables are independent. Mutual information can capture arbitrary statistical dependence.

\begin{figure}[ht]
    \centering
    \begin{minipage}[t]{0.48\columnwidth}
        \centering
        \stackinset{l}{0.1em}{t}{0.1em}{\small\textbf{(a)}}{
            \includegraphics[width=1\linewidth]{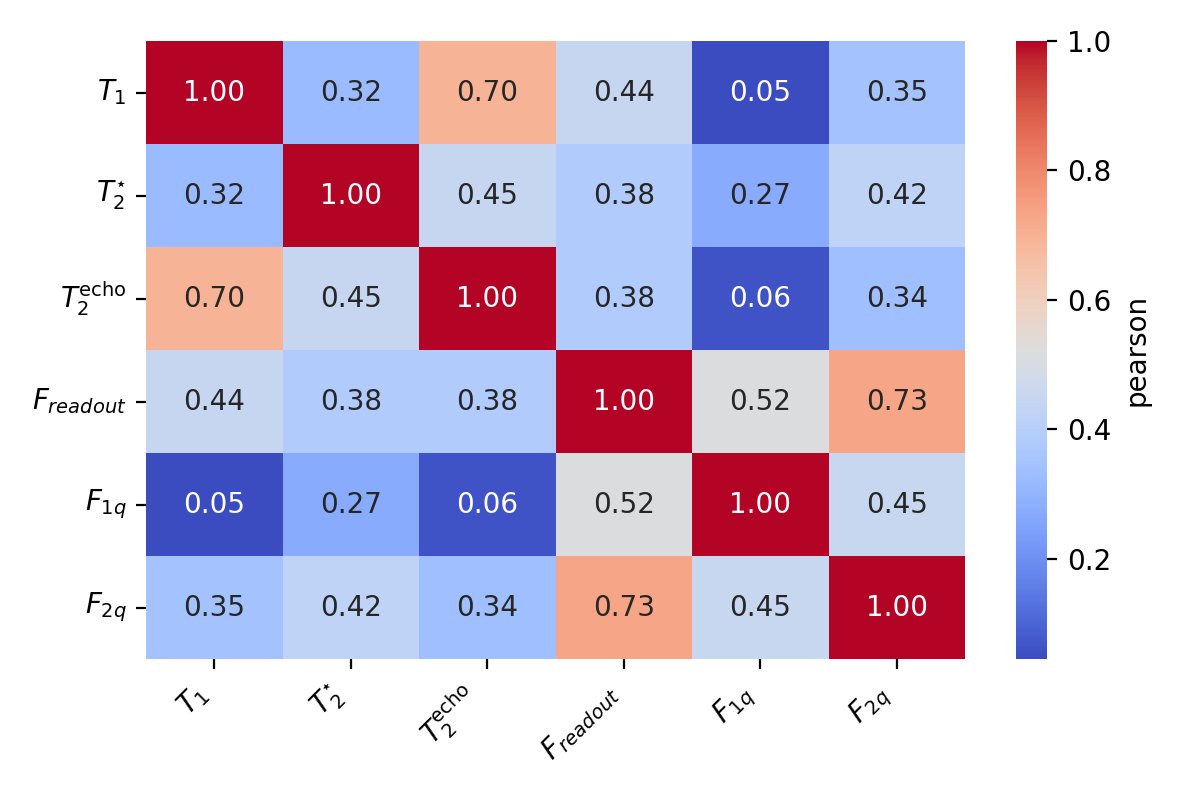}
	}
        \label{fig:corr_pearson}
    \end{minipage}
    \hfill 
    \begin{minipage}[t]{0.48\columnwidth}
        \centering
        \stackinset{l}{0.1em}{t}{0.1em}{\small\textbf{(b)}}{
            \includegraphics[width=1\linewidth]{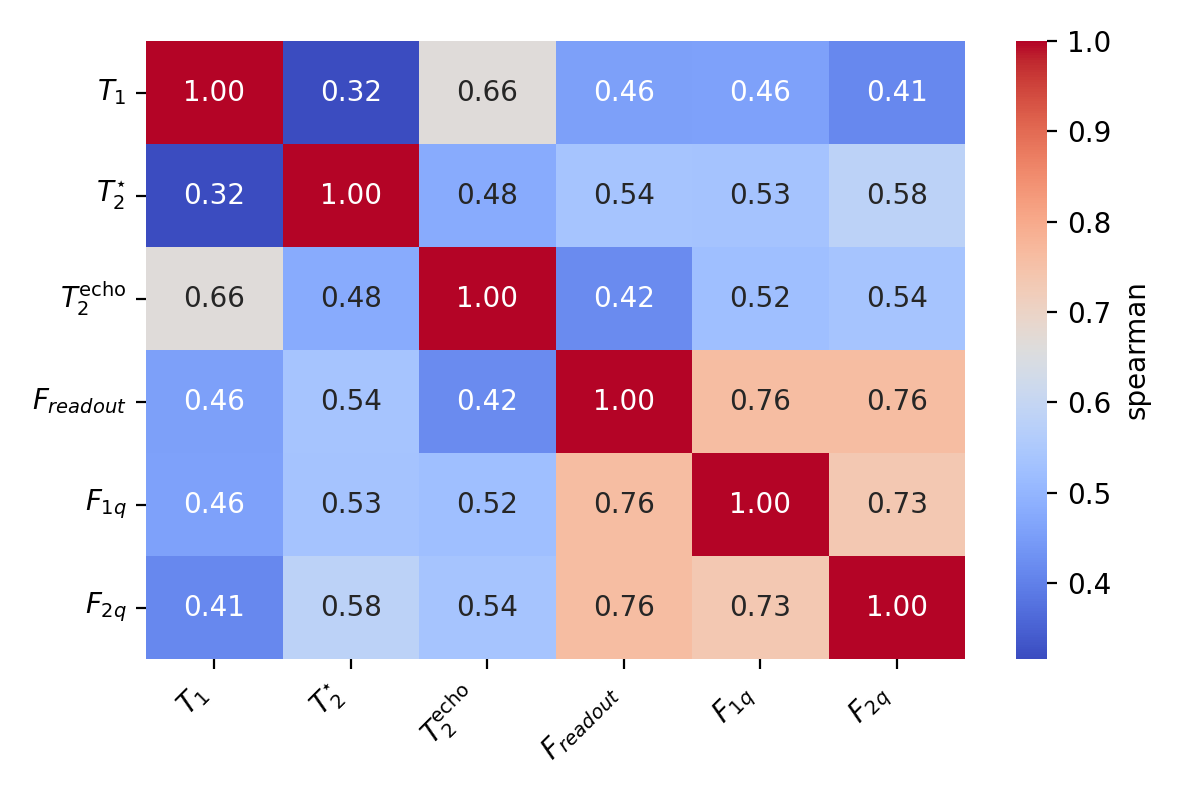}
        }
        \label{fig:corr_spearman}
    \end{minipage}
\hfill
    \begin{minipage}[t]{0.48\columnwidth}
        \centering
        \stackinset{l}{0.1em}{t}{0.1em}{\small\textbf{(c)}}{
            \includegraphics[width=1\linewidth]{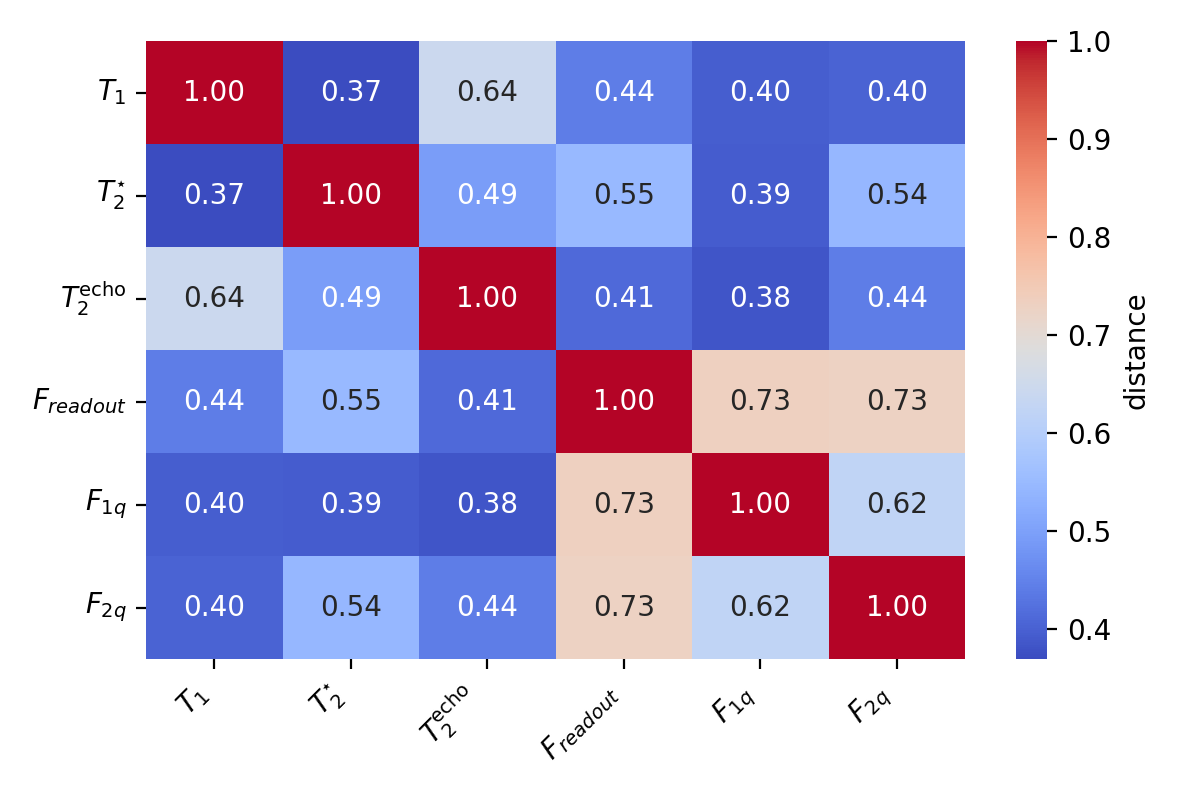}
	}
        \label{fig:corr_distance}
    \end{minipage}
    \hfill 
    \begin{minipage}[t]{0.48\columnwidth}
        \centering
        \stackinset{l}{0.1em}{t}{0.1em}{\small\textbf{(d)}}{
            \includegraphics[width=1\linewidth]{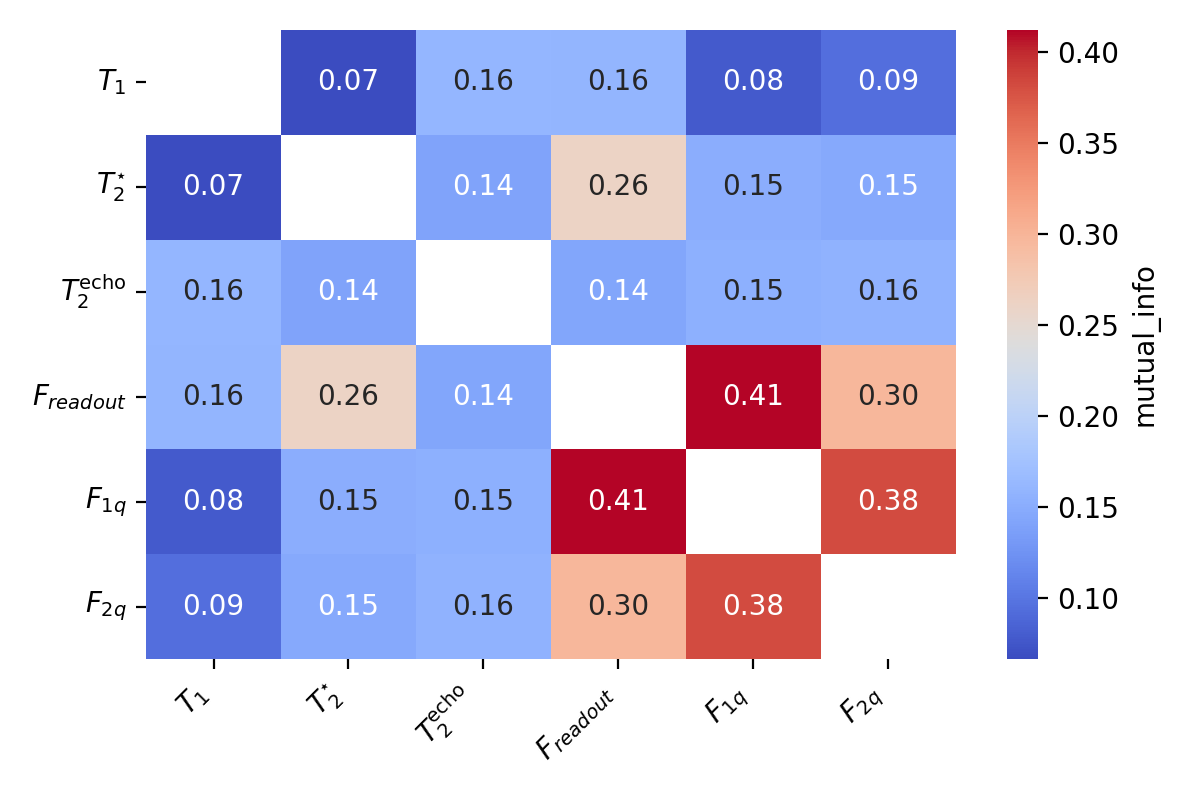}
        }
        \label{fig:corr_mutual}
    \end{minipage}

    \caption{Metric-metric correlation of daily-mean quality metrics over a time window ($80$ days) from a cool-down. (a) Pearson; (b) Spearman; (c) Distance correlation; (d) Mutual information. Fidelity metrics form a strongly-coupled block and coherence times form a second block. Cross-block couplings are weak to moderate.}
    \label{fig:metric_metric_corr}
\end{figure}

We calculate pairwise correlations between daily mean time series of all main quality metrics (over 80 days after cool-down, see Fig. \ref{fig:metric_metric_corr}). All metrics are first standardized (zero mean, unit variance) so that differences in physical units do not bias the estimates. Remarkably, all four correlation methods yield consistent patterns. Fidelity metrics (readout, single-qubit, two-qubit) form a tightly coupled cluster. This means that one fidelity drops, the others are highly likely to go down as well. This dependence reflects the shared sensitivity of these metrics to the global factors, such as temperature drift or control electronics. Coherence times also show strong internal correlations, especially between $T_1$ and $T^{\rm echo}_2$. This reflects the common underlying noise resources. Fidelity and coherence times are only weakly correlated shown in the figure.

These findings have direct impact on both daily operations and long-term maintenance. Since fidelity metrics are strongly correlated, monitoring any of them may be sufficient for the whole system health. This can simplify automated anomaly detection and reduce computational burden. The cross-metric correlation matrices can also help design early alert systems. For instance, if $T_1$ and $T^{\rm echo}_2$ simultaneously decrease, it is likely that a broader failure appears and a recalibration need to be triggered. Thirdly, in daily calibration data monitoring, distance correlation can be used as the core method, supplemented by Spearman correlation to verify monotonic relationships. When the sample size is large enough, mutual information can be helpful to explore complex dependencies.

The above methods can be applied to per-qubit cross-metric correlations or lagged pairs as well. Due to the limited space, it is not shown here.
\subsection{Unsupervised clustering}
Heatmaps, variances, and daily-mean correlations reveal average patterns and overall trends in quantum device metrics drifting collectively over time. In practice, individual qubits exhibit significantly different behaviours, such as $T_1$/$T_2$, gate/readout error rate and characteristic drift behaviours can differ by orders of magnitudes across different qubits. These differences are influenced by qubit position, fabrication variations and shared environment or couplings in the chip's connectivity graph.

\begin{figure}[ht]
    \centering
    \begin{minipage}[t]{0.48\columnwidth}
        \centering
        \stackinset{l}{0.1em}{t}{0.1em}{\small\textbf{(a)}}{
            \includegraphics[width=1\linewidth]{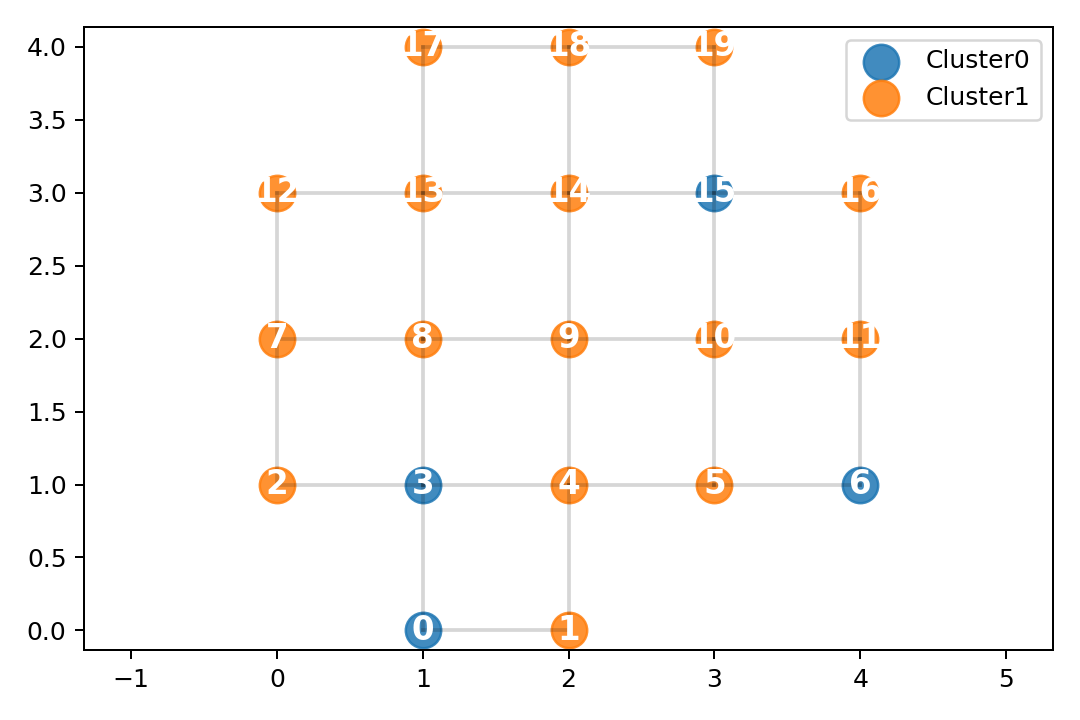}
	}
        \label{fig:node2vec}
    \end{minipage}
    \hfill 
    \begin{minipage}[t]{0.48\columnwidth}
        \centering
        \stackinset{l}{0.1em}{t}{0.1em}{\small\textbf{(b)}}{
            \includegraphics[width=1\linewidth]{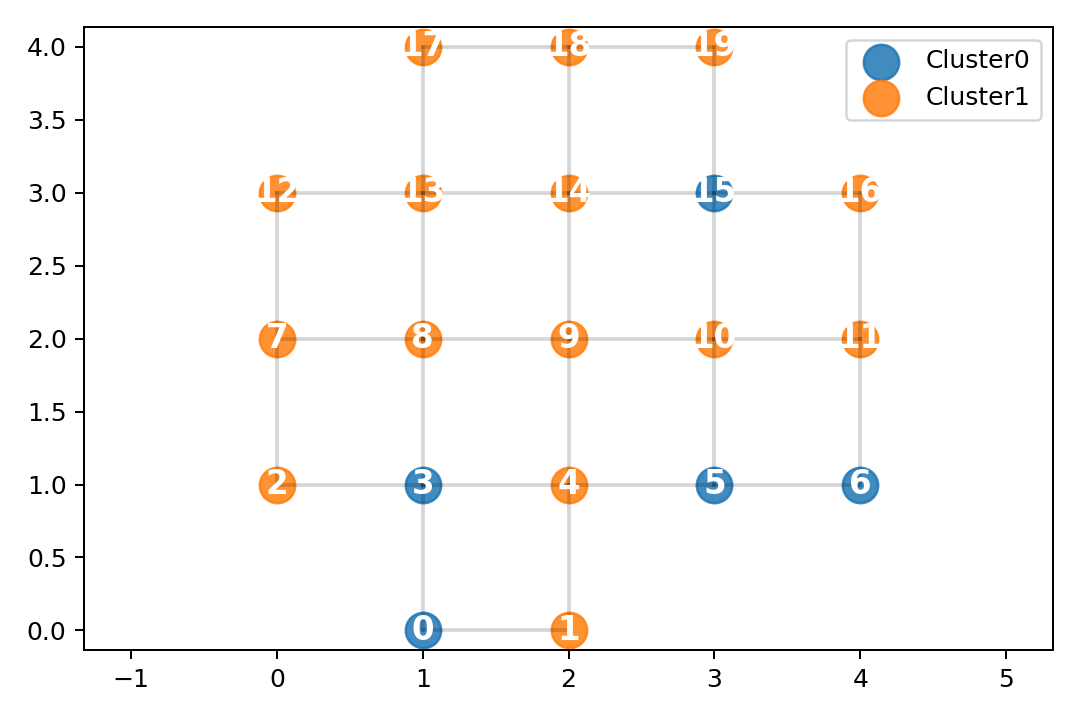}
        }
        \label{fig:gmm}
    \end{minipage}
 \caption{Clustering of qubits by (a) Node2Vec+KMeans and (b) GMM. The results by KMeans and Spectral give the same as (a).}
    \label{fig:clustering}
\end{figure}

To systematically uncover these patterns of heterogeneity, we employ unsupervised clustering on the calibration dataset. We combine each qubit's temporal feature vector (e.g., daily means of 6 dimensional vector) with their topological connectivity (via Node2Vec graph embedding\cite{10.1145/2939672.2939754}). The optimal number of clusters are determined by maximizing the silhouette score. Each resulting cluster represents a group of qubits that behave similarly over a calibration window, potentially indicating local defects, common noise environments or couplings. We use a variety of clustering algorithms: KMeans, GMM, Spectral, and Node2Vec+KMeans\cite{hastie2009elements, DBLP:journals/sac/Luxburg07, 10.1145/2939672.2939754}. The resulting clusters consistently separate the qubit population into stable and noisy families, as shown in Fig. \ref{fig:clustering}. Most qubits form a large, stable cluster, while a smaller group exhibits lower performance. This clustering is robust across different algorithms, confirming that the observed groups are intrinsic properties of the device rather than artifacts of a specific method.

Clustering provides actionable directions for calibration management. Clusters that consistently contain noisy qubits need prioritized recalibration, diagnosis, or exclusion from critical circuit mapping. Conversely, large stable clusters can be prioritized for deep or error-sensitive circuits. Integrating clustering into HPCQC workflows enables automated, data-driven decisions for circuit scheduling and resource allocation.

\section{Use Cases}\label{sec:use}
Building on our long-term analysis and unsupervised clustering, we now turn to operational validation using real quantum circuits. While daily calibration provides a broad overview of device health, quantum processors can experience significant drift on shorter timescales, e.g., within minutes or hours. To capture these finer variations and to verify our clustering results in practice, we execute short quantum circuits, such as Hadamard gate circuits and GHZ states, every hour across all qubits and pairs. These frequent circuit benchmarks have a dual purpose: they act as sensitive probes for detecting performance fluctuations missed by daily sampling, and they provide the experiments to validate our clustering analysis.

\begin{figure}[ht]
    \centering
    \begin{minipage}[t]{0.48\columnwidth}
        \centering
        \stackinset{r}{0.1em}{t}{0.1em}{\small\textbf{(a)}}{
            \includegraphics[width=1\linewidth]{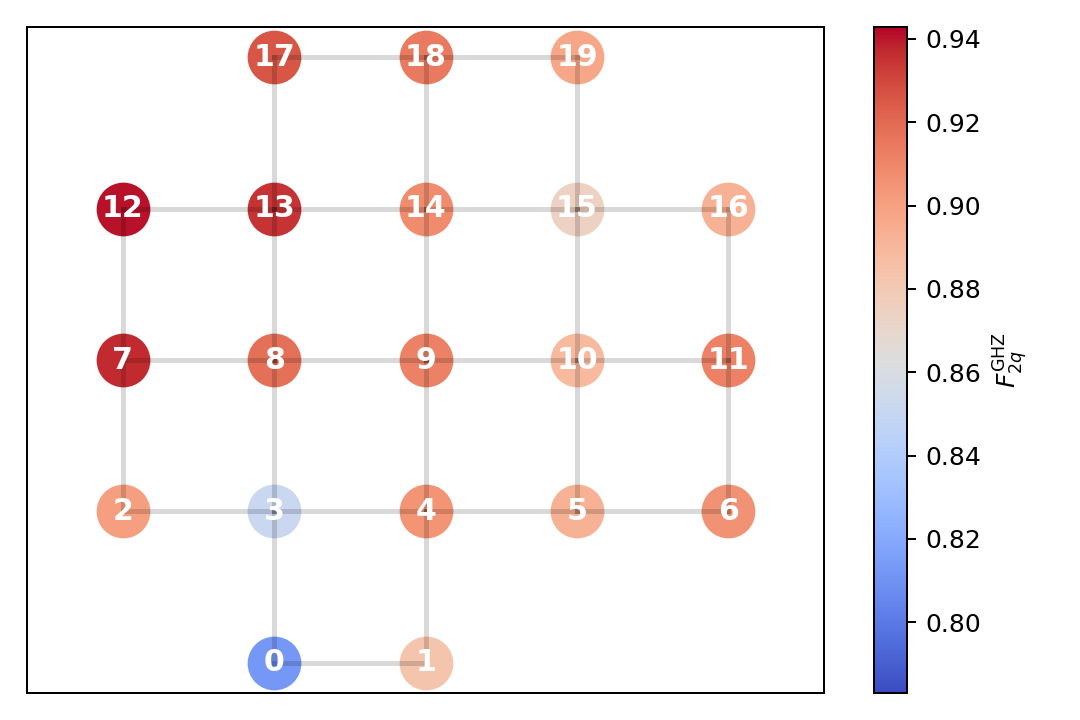}
	}
        \label{fig:ghz2}
    \end{minipage}
    \hfill 
    \begin{minipage}[t]{0.48\columnwidth}
        \centering
        \stackinset{l}{0.1em}{t}{0.1em}{\small\textbf{(b)}}{
            \includegraphics[width=1\linewidth]{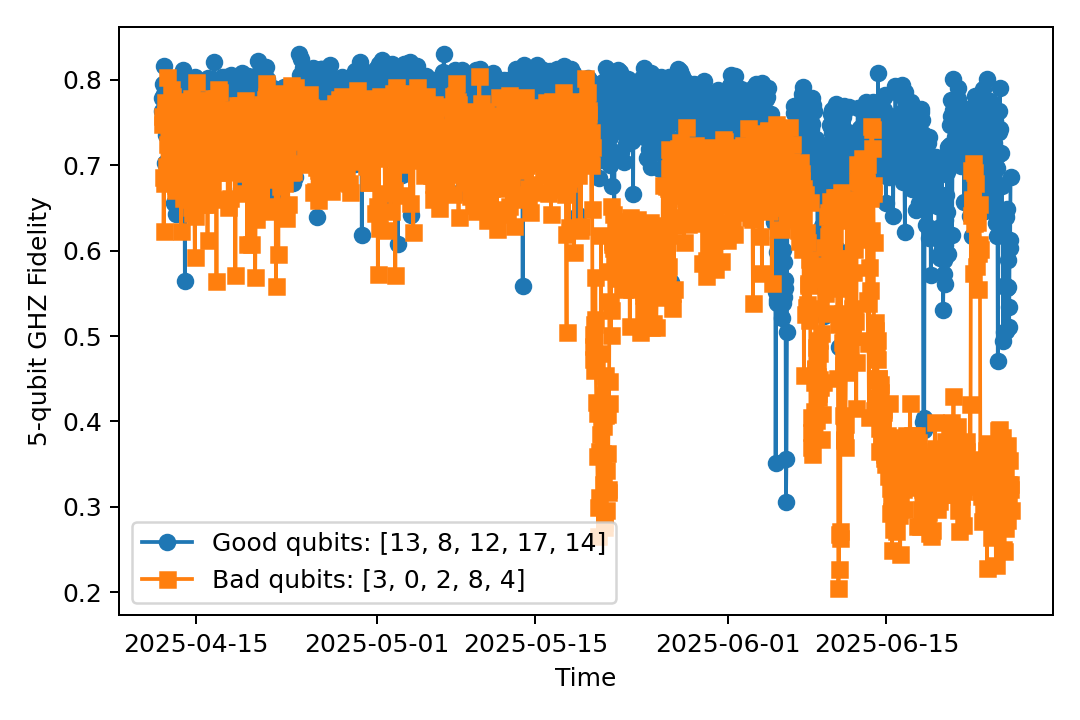}
        }
        \label{fig:ghz5}
    \end{minipage}
 \caption{(a) Heatmap of 2-qubit GHZ fidelity of each pair, that is assigned to its paired qubits. (b) 5-qubit GHZ state evolution formed by good qubits $\{13,8,12,17,14\}$ (top left) and bad qubits $\{3,0,2,8,4\}$ (bottom left). The mean fidelity of good circuit is $0.74\pm 0.05$, while the mean of bad circuit is $0.63\pm0.14$. }
    \label{fig:ghz_verify}
\end{figure}

Fig. \ref{fig:ghz_verify} (a) shows a heatmap of the mean fidelity of 2-qubit GHZ circuits on all coupled pairs, over the measurement window and mapped to the physical layout. This pattern closely matches the clusters identified by our earlier unsupervised learning methods. Regions of high GHZ fidelity correspond to clusters of stable qubits, where regions with lower fidelity correspond to identified noisy clusters.

To further quantify the impact of clustering on operation, we compare the fidelity evolution of two representative 5-qubit GHZ circuits over time. One is formed by qubits from the stable cluster, and one is from the noisy cluster, see Fig. \ref{fig:ghz_verify} (b). The stable ("good") cluster circuit maintains consistently higher fidelity ($0.74 \pm 0.05$), while the noisy (“bad”) cluster circuit shows both lower mean fidelity ($0.63 \pm 0.14$) and increased variability. This demonstrates that circuits mapped to robust clusters indeed yield more reliable experimental outcomes.

These use cases demonstrate the value of data-driven qubit selection and mapping for practical hardware operation. By integrating validation with long-term calibration monitoring, operators (in HPC centers) can make more informed decisions on recalibration, resource allocation and maintenance scheduling. This is particularly relevant for large-scale, hybrid quantum-classical workflows in HPC centers\cite{11017506}.

\section{Conclusion}\label{sec:conclusion}
In summary, we have presented in this work a detailed data-driven framework for long-term calibration analysis and predictive health monitoring of a superconducting quantum processor. By using over 250 days of calibration data from a 20-qubit NISQ device, we have systematically analyzed the temporal evolution and cross-metric correlations of key quality metrics, including $T_1$, $T_2$ and gate/readout fidelities. Using different unsupervised learning methods combining metric features and QPU topology we have robustly clustered qubits into stable and noisy sets. That provides indicators for targeted recalibration and circuit mapping. Our framework has been validated via direct experiments, such as hourly Hadamard and multi-qubit GHZ fidelities, and the results have shown strong agreement between predicted good clusters and actual operational performance. Our methods are practical for device health monitoring and scheduling.

In future work we will expand this approach in several directions. In HPC integration of quantum systems it would be interesting to examine whether the HPC environment leads to different failure/performance degradation patterns than in less demanding setups. We will thus integrate environmental factors, e.g., temperature, magnetic field, vibration, etc. to the framework. We will also extend predictive models to spatio-temporal neural network and discrete Bayesian network, and evaluate generalization to other quantum hardware platforms. As our HPC center is preparing to deploy 54-qubit and 150-qubit chips in the near future, we also plan to extend our methodology to larger quantum systems. This will enable us to develop practical scaling laws of calibration metrics within supercomputing environments.

\section*{Acknowledgment}
We thank the LRZ QCT-SL team for granting us access to the operational data of Q-Exa system and the LRZ GM team for continued support. Special thanks to the IQM team for collaborative efforts in setting up the full calibration and recalibration routines. The work was funded by the BMBF Q-Exa project, Euro-Q-Exa project, MUNIQC-SC project, and the Munich Quantum Valley (MQV), which is supported by the Bavarian State
Government with funds from the Hightech Agenda Bayern.
\bibliographystyle{IEEEtran}

\bibliography{saqs-base}

\end{document}